\documentclass[journal]{IEEEtran}
\usepackage{times,amsfonts,amssymb,amsmath,setspace}
\usepackage{color}
\usepackage{graphicx,multirow}
\usepackage{cite}
\usepackage{url}
\usepackage{mathrsfs}
\usepackage{float}
\usepackage{lipsum}
\usepackage{multicol}
\usepackage{mathtools} 
\usepackage{graphicx,hyperref,bm,subfigure,comment}

\DeclareMathAlphabet\mathbfcal{OMS}{cmsy}{b}{n}
\newcommand{\insertfig}[4]{
\begin{figure}[ht]
\centerline{\includegraphics[width=#1\columnwidth]{#2.eps}}
\vspace{-3ex}
\caption{#3}\label{#4}\end{figure}}

\DeclareMathAlphabet{\mathsfbf}{OT1}{cmss}{sbc}{n}

\newtheorem{proposition}{Proposition}[section]

\newcommand{\EE}{\mathbb{E}} 
\newcommand{\PP}{\mathbb{P}} 

\newcommand{\ee}{{\rm e}}

\newcommand{\av}{{\bf a}}

\newcommand{\hv}{{\bf h}}

\newcommand{\mv}{{\bf m}}
\newcommand{\nv}{{\bf n}}

\newcommand{\tv}{{\bf t}}

\newcommand{\vv}{{\bf v}}

\newcommand{\Am}{{\bf A}}

\newcommand{\Dm}{{\bf D}}

\newcommand{\Gm}{{\bf G}}

\newcommand{\Mm}{{\bf M}}

\newcommand{\Bb}{\mathbfcal B}
\newcommand{\Fb}{\mathbfcal F}

\newcommand{\Ac}{{\mathcal A}}
\newcommand{\Bc}{{\mathcal B}}
\newcommand{\Cc}{{\mathcal C}}

\newcommand{\Gc}{{\mathcal G}}
\newcommand{\Hc}{{\mathcal H}}

\newcommand{\Lc}{{\mathcal L}}

\newcommand{\Oc}{{\mathcal O}}

\newcommand{\alphav}{\boldsymbol{\alpha}}

\newcommand{\tauv}{\boldsymbol{\tau}}

\newcommand{\sgn}{\text{sgn}}

\def\ben{\begin{enumerate}}
\def\beq{\begin{equation}}
\def\beqa{\begin{eqnarray}}
\def\bit{\begin{itemize}}
\def\een{\end{enumerate}}
\def\eeq{\end{equation}}
\def\eeqa{\end{eqnarray}}
\def\eit{\end{itemize}}

\def\non{\nonumber\\}
\def\argmax{\mathop{\mathrm{arg~max}}\limits}

\begin{document}

\title{The Importance of Worker Reputation Information \\ 
in {\color{black} Microtask-Based} Crowd Work Systems} 
\author{Alberto Tarable, Alessandro Nordio, Emilio Leonardi, Marco Ajmone Marsan
\thanks{A. Tarable and A. Nordio are with CNR-IEIIT, Torino, Italy. E-mail: \{alberto.tarable;  alessandro.nordio\}{@}ieiit.cnr.it.}
\thanks{E. Leonardi and M. Ajmone Marsan are with the Department of Electronic and Telecommunications, Politecnico di Torino, Torino, Italy and are also Research Associates with CNR-IEIIT, Torino, Italy. E-mail: \{emilio.leonardi, marco.ajmone\}{@}polito.it.} 
\thanks{M. Ajmone Marsan is a part time Research Professor at IMDEA Networks Institute in Leganes, Madrid, Spain.}
}
\maketitle

\begin{abstract} 
This paper presents the first systematic investigation of the potential
performance gains for crowd work systems, deriving from available information at
the requester about individual worker reputation. In particular, we first
formalize the optimal task assignment problem when workers' reputation estimates
are available, as the maximization of a monotone (submodular) function subject
to Matroid constraints. Then, being the optimal problem NP-hard, we propose a
simple but efficient greedy heuristic task allocation algorithm.  We also
propose a simple ``maximum a-posteriori'' decision rule and a decision algorithm
based on message passing. Finally, we test and compare different solutions,
showing that system performance can greatly benefit from information about
workers' reputation.  Our main findings are that: i) even largely inaccurate
estimates of workers' reputation can be effectively exploited in the task
assignment to greatly improve system performance; ii) the performance of the
maximum a-posteriori decision rule quickly degrades as worker reputation
estimates become inaccurate; iii) when workers' reputation estimates are
significantly inaccurate, the best performance can be obtained by combining our
proposed task assignment algorithm with the message-passing decision algorithm.
\end{abstract}


\section{Introduction}

Crowd work is a term often adopted to identify networked systems that can be
used for the solution of a wide range of complex problems by integrating a large
number of human and/or computer efforts \cite{asurveyofcrowdsourcingsystems}.
Alternative terms, each one carrying its own specific nuance, to identify
similar types of systems are: collective intelligence, human computation,
master-worker computing, volunteer computing, serious games, voting problems,
peer production, citizen science (and others).  {\color{black} An entire host of
  general-purpose or specialized online platforms, such as information-sharing
  platforms for recommendations (e.g., Tripadvisor, Amazon), co-creation systems
  (e.g., Wikipedia, Gnu project), social-purpose communities for urban mobility
  (e.g., Waze), microtask-based crowd work systems, etc., can be defined under
  these terms.

In this paper, we specialize to microtask-based crowd work systems.}  The key
characteristic of these systems is that a {\em requester} structures his problem
in a set of {\em tasks}, and then assigns tasks to {\em workers} that provide
{\em answers}, which are then used to determine the correct task {\em solution}
through a {\em decision} rule.  A well-known example of such systems is Amazon
Mechanical Turk {\color{black} (MTurk)}, which allows the employment of large
numbers of {\color{black} low-wage} workers for tasks requiring human
intelligence (HIT -- Human Intelligence Tasks). Examples of HIT are image
classification, annotation, rating and recommendation, speech labeling,
proofreading, etc. In the Amazon Mechanical Turk, the workload submitted by the
requester is partitioned into several {\color{black} microtasks}, with a simple
and strictly specified structure, which are then assigned to (human)
workers. Since task execution is typically tedious, and the economic reward for
workers is pretty small, workers are not 100\% reliable, in the sense that they
may provide incorrect answers. Hence, {\color{black} in most practical cases},
the same task is assigned in parallel (replicated) to several workers, and then
a majority decision rule is applied to their answers. A natural trade-off
between reliability of the decision and cost arises; indeed, by increasing the
replication factor of every task, we generally increase the reliability degree
of the final decision about the task solution, but we necessarily incur higher
costs (or, for a given fixed cost, we obtain a lower task throughput).  Although
the pool of workers in crowd work systems is normally large, it can be
abstracted as a finite set of shared resources, so that the allocation of tasks
to workers (or, equivalently, of workers to tasks) is of key relevance to the
system performance.
Some believe that {\color{black} microtask-based} crowd work systems will provide
a significant new type of work organization paradigm, and will employ
{\color{black} ever increasing~\cite{Difallah}} numbers of workers in the future,
provided that the main challenges in this new type of organizations are
correctly solved. In \cite{thefutureofcrowdwork} the authors identify a dozen
such challenges, including i) workflow definition and hierarchy, ii) task
assignment, iii) real-time response, iv) quality control and reputation.
{\color{black} All these aspects can represent an interesting research subject
  and some of them have already stimulated a large bulk of literature, as it
  will be detailed in the next subsection. However, this paper deals mainly with
  task assignment and with the quantitative assessment of the gain (in terms of
  increased decision reliability for a given cost) that a coarse knowledge of
  worker quality can offer. Indirectly, thus, we deal also with worker
  reputation, although we do not study mechanisms through which reputation is
  built upon time.  Indeed, we consider a one-shot approach in which the
  requester has to assign a bunch of tasks to a pool of workers that are
  statically divided into \emph{classes} according to their probabilities of
  answering correctly. We highlight that the way this division into classes is
  built is out of the scope of this paper, although we will analyze the effect
  of errors in this classification on the decision reliability.}
 
\subsection{Previous work}

{\color{black} In current online platforms, task assignment is either implemented
  through a simple first-come/first-served rule, or according to more
  sophisticated approaches. In MTurk, the requester can specify the number of
  workers to be assigned to each task. MTurk also gives requesters the
  possibility of dismissing low-quality answers, so that each experienced worker
  is characterized by an approval rating. As a consequence, the requester is
  also allowed to prescribe a given qualification level for workers to be able
  to access her tasks. An analysis of the correlation between MTurk approval
  rating and worker quality is performed in~\cite{Peer}. In the scientific
  community, the task assignment in crowdsourcing systems has recently been
  formalized~\cite{Devavrat, shah2, shah3, moderators} as a resource allocation
  problem, under the assumption that both tasks and workers are
  indistinguishable. On the worker side, this assumption is motivated by the
  fact that the implementation of reputation-tracing mechanisms for workers may
  be challenging, because the workers' pool is typically large and highly
  volatile. A step ahead has been recently made in~\cite{adaptive}, which
  proposes an adaptive online algorithm to assign an appropriate number of
  workers to every task, so as to meet a prefixed constraint on the problem
  solution reliability. Like in this paper, in~\cite{adaptive} workers are
  partitioned in different classes, with workers within each class meeting a
  specified reliability index. However, unlike this paper, the allocation
  algorithm of~\cite{adaptive} is adaptive, i.e., it is based on previous
  answers on the same set of microtasks: an assumption that, although certainly
  interesting, implies a time-consuming overall process of task
  accomplishment. The same adaptive approach is followed in~\cite{Zhang}, where
  a bandit-based algorithm is adopted to allocate heterogeneous tasks to workers
  with task-dependent skills. Given a pool of $n$ questions,~\cite{Zheng}
  investigates how $k$ questions therefrom should be assigned to a worker.

Most real-world crowdsourcing systems typically implement a majority-based
decision rule to obtain task solutions. In the last few years, in the scientific
literature, smarter decision rules have been proposed to improve the performance
of crowdsourcing systems, for the case where no a-priori information about
workers' reputation is available (i.e. workers are a-priori indistinguishable)
while tasks are homogeneously difficult~\cite{Devavrat, shah2, shah3,
  moderators, Bachrach, Raykar}. Essentially the same decision strategy was
proposed in~\cite{Devavrat,shah2} and~\cite{moderators} for the case in which
tasks require binary answers, and then recently extended in~\cite{shah3} and,
independently, in~\cite{Lee}, to the case in which tasks require generic
multiple-choice answers. In~\cite{shah2,shah3} it is shown that the improved
decision rule can be efficiently implemented employing a message-passing
technique.  In~\cite{Bachrach}, an integrated estimation-allocation approach has
been pursued with Bayesian inference and entropy reduction as utility
function. Different methodologies based on the Expectation-Maximization
algorithm have been proposed in~\cite{Whitehill,Raykar}. All these algorithms
exploit existing redundancy and correlation in the pattern of answers returned
from workers to infer an a-posteriori reliability estimate for every worker. The
derived estimates are then used to properly weigh workers' answers. When there
is a-priori information about workers' reliability, to the best of our
knowledge, the only decision rule proposed in the literature is weighted
average, like, e.g., in~\cite{Zhang}.

There is a wide literature about workers' motivation and reputation. Regarding
motivation, many studies report experiments conducted on MTurk as a paradigm of
microtask-based crowd work platform. Classical studies of offline work reveal
that workers try to understand which activities are better rewarded and tend to
prefer those, virtually excluding others~\cite{Kerr}. However, in the context of
crowdsourcing systems, mixed results have been shown about the effect of
economic incentives on the quality of workers' outputs~\cite{Chandler, Kittur4,
  Lewis, Mason, Rogstadius, Shaw}.  These studies highlight the importance of
intrinsically non-economical motivations such as the feeling of contributing
towards a greater good, non-financial awards and recognitions, accomplishing
tasks that appear meaningful. An attempt of a systematic model of crowdsourcing
workers with respect to financial incentives is proposed by~\cite{Ho}, in which
experiments carried out on MTurk reveal that a monetary bonus is effective when
it is large enough compared to the base payment, while it saves money with
respect to a generalized increase of the latter, for the same answers'
quality. Reputation mechanisms are an important tool for crowdsourcing systems
amd are active already in many online platforms. For example, UpWork implements
a feedback system, which is bidirectional (workers vote requesters and vice
versa). While Upwork distributes larger and more complex tasks, in
microtask-based platforms, feedback is more limited. As already said, MTurk
characterizes the workers with an approval rating. In the scientific literature,
examples of algorithms that incorporate auditing processes in a sequence of task
assignments for worker reputation assessment can be found in
\cite{antonio1,antonio2,antonio3,econ1,econ2,KDD09,ZSD10}. In~\cite{Christoforou},
a dynamical reputation mechanisms is devised for a crowd composed of three types
of workers: altruistic, malicious and rational. It is shown
in~\cite{Christoforou} that, with a proper dimensioning of the financial rewards
and penalties, the probability of an audit (where the requester herself executes
the task) tends to zero.  More in general,~\cite{Gadiraju} studies the impact of
malicious workers on the performance of crowdsourcing systems.  It is recognized
in~\cite{Kokkodis} that the reputation of each worker must differentiate
according to different types of task. In~\cite{Fan}, a task similarity graph is
used to infer workers’ reliabilities for open tasks based on their performance
on completed tasks. An aspect closely related to reputation is online workers'
quality control. Some systems, such as UpWork Worker Diary, make available to
the requester periodic snapshots of workers' computer screens, to increase
visibility on how the employed workers are behaving. The impact of the so-called
attention-check questions (trick questions inserted in a task in order to test
the worker's attention) is analyzed in~\cite{Peer}, where it is concluded that
such questions are useful only for low-reliability workers and may be
counter-productive for high-reliability workers.  Workers' quality control can
be partly automated, and made more effective by employing machine-learning
techniques like reinforcement learning~\cite{Rzeszo,Rzeszo2}. Finally, regarding
workers' organization,~\cite{Buettner} presents a comprehensive literature
survey on human resource management in crowdsourcing systems.}

\subsection{Our contribution}
 
Task assignment and
reputation are central to this paper, where we discuss optimal task assignment with approximate
information about the quality of answers generated by workers (with the term ``worker reputation'' we
generally mean the worker earnestness, i.e., the credibility of a worker's answer for a given task,
which we will quantify with an error probability). Our optimization aims at minimizing the
probability of an incorrect task solution for a maximum number of tasks assigned to workers, thus
providing an upper bound to delay and a lower bound on throughput. A dual version of our
optimization is possible, by maximizing throughput (or minimizing delay) under an error probability
constraint. Like in most analyses of crowd work systems, we assume no interdependence among
tasks, but the definition of workflows and hierarchies is an obvious next step. Both these issues
(the dual problem and the interdependence among tasks) are left for further work.

The goal of this paper is to provide the first systematic analysis of the
potential benefits deriving from some form of a-priori knowledge about the
reputation of workers, extending the results of our preliminary
work~\cite{Infocom2015}.  With this goal in mind, first we define and analyze
the task assignment problem when workers' reputation estimates are
available. {\color{black} In particular, we suppose that available workers are
  divided into classes, each of which represents a different reliability
  level. Under this hypothesis,} we show that in some cases, the task assignment
problem can be formalized as the maximization of a monotone submodular function
subject to Matroid constraints. A greedy algorithm with performance guarantees
is then devised.  In addition, we propose a simple ``maximum a-posteriori``
(MAP) decision rule, which is well known to be optimal when perfect estimates of
workers' reputation are available.  Moreover, we introduce a message-passing
decision algorithm, which is able to encompass a-priori information about
workers' reputation, thus improving upon the one described in \cite{shah2}.
Finally, our proposed approach is tested in several scenarios, and compared to
previous proposals.

Our main findings are:
\begin{itemize}
\item even largely inaccurate estimates of workers' reputation can be effectively exploited in the
  task assignment to greatly improve system performance;
\item the performance of the maximum a-posteriori decision rule quickly degrades as worker
  reputation estimates become inaccurate;
\item when workers' reputation estimates are significantly inaccurate, the best performance can be
  obtained by combining our proposed task assignment algorithm with the message-passing decision algorithm presented in this paper;
\item {\color{black} there is no need for a large number of refined classes, i.e., a coarse quantization of individual reputations already achieves most of the related gain.}  
\end{itemize}


\section{System Assumptions}
\label{sec:SA}

We consider $T$ binary tasks $\theta_1, \theta_2, \dots, \theta_T$, whose outcomes can be
represented by i.i.d. uniform random variables (RV's) $\tau_1, \tau_2, \dots, \tau_T$ over $\{\pm
1\}$, i.e., $ \PP \{\tau_t = \pm 1\} = \frac1{2}$, $t = 1,\dots, T$. In order to obtain a reliable
estimate of task outcomes, a requester assigns tasks to workers selected from a given population of
size $W$, by querying each worker $\omega_w$, $w = 1,\dots, W$ a subset of tasks.

Each worker is modeled as a binary symmetric channel (BSC)
\cite[p. 8]{Cover}. This means that worker $\omega_w$, if queried about task
$\theta_t$, provides a wrong answer with probability $p_{t w}${\color{black},
  $p_{t w} \in [0,1/2]$, and a correct answer with probability $1-p_{t w}$. The
  error probabilities $p_{t w}$ are taken to be time-invariant and generally
  unknown to the requester.

\medskip
\noindent {\bf Remark 1} In practice, $p_{t w}$ can be
  estimated by analyzing the workers' performance during previous task
  assignments. However how to estimate $p_{t w}$ is out of the scope of this
  work.

\medskip
\noindent {\bf Remark 2} We assume that the task allocation process works in
one-shot.  More precisely, at time $t$ the allocation algorithm submits all
tasks to workers on the basis of the reputation they have at that
time. Therefore, possible time variations of the workers' reputation do not
affect task allocation. Also, tasks are assumed to be evaluated by workers in a
short amount of time. Therefore workers' error probabilities, $p_{tw}$, are not
expected to significantly vary during the process.  An analysis of the system
performance in the presence of time variations of workers' reputations would
require models and algorithms for building reputation on the basis of previously
assigned tasks.  However, our scope is not to investigate how these reputations
can be built, rather to assess how reputation can be exploited by task
allocation and decision algorithms.
\medskip

  By making these assumptions, we avoid modeling the
  workers' behaviour as driven by latent motivations, as in,
  e.g.,~\cite{Christoforou}. In particular, we do not deal with malicious
  workers (for which $p_{t w} = 1$). As a matter of fact, a worker that always
  outputs the wrong answer provides as much information to the aware requester
  as a worker that answers correctly all the times. We also assume that $p_{t
    w}$ depends on both the worker and the task, a fact that} reflects the
realistic consideration that tasks may have different levels of difficulty, that
workers may have different levels of accuracy, and may be more skilled in some
tasks than in others{\color{black}~\cite{Kokkodis}}.

{\color{black} Similarly to \cite{adaptive}}, we assume in this paper that,
thanks to a-priori information, the requester can group workers into
\emph{classes}, each one composed of workers with similar accuracy and
skills. In practical crowd work systems, where workers are identified through
authentication, such a-priori information can be obtained {\color{black} at no
  cost} by observing the results of previous task assignments{\color{black}: this
  is the case of the approval rating in MTurk, for example}. More precisely, we
suppose that each worker belongs to one of $K$ classes, $\Cc_1, \Cc_2, \dots,
\Cc_K$, and that each class is characterized, for each task, by a different
   {\color{black} representative} error probability, known to the requester. Let
   $\pi_{t k}$ be the {\color{black} representative} error probability for class
   $\Cc_k$ and task $\theta_t$, $k = 1,\dots, K$, $t = 1,\dots, T$.
   {\color{black} In practice, classes may derive from a quantization of
     estimated individual error probabilities. The reason of dealing with
     classes, instead of individuals, stems from the fact that $p_{t w}$ is
     estimated heuristically and thus it is affected by inaccuracy. Because of
     that, sacrificing precision should not entail a major performance loss,
     while it simplifies the task allocation phase. This intuition will be
     confirmed in Section~\ref{sec:results}.}

{\color{black} Most times, in the following, we will deal with a case where $\pi_{t k}$ is the {\color{black} \emph{average}} error probability of workers belonging to class $k$. This allows to abstract from the practical way in which classes are built.}
In particular our class
characterization encompasses two extreme scenarios:

\begin{itemize}
\item full knowledge about the reliability of workers, i.e., each worker belonging to class $\Cc_k$
  has error probability for task $\theta_t$ deterministically equal to $\pi_{t k}$, and
\item a hammer-spammer (HS) model~\cite{Devavrat}, in which perfectly reliable and completely
  unreliable users coexists within the same class.  A fraction $2\pi_{t k}$ of workers in class
  $\Cc_k$, when queried about task $\theta_t$, has error probability equal to $\frac1{2}$ (the
  spammers), while the remaining workers have error probability equal to zero (the hammers). {\color{black} Note that this is an artificial scenario, where the variance within a single class is pushed to the limit, thus allowing to test the robustness of our task assignment algorithm to a very unfavorable class composition}.
\end{itemize}

Suppose that class $\Cc_k$ contains a total of $W_k$ workers, with $W = \sum_{k=1}^K W_k$.  The
first duty the requester has to carry out is the assignment of tasks to workers. We impose the
following two constraints on possible assignments:
\begin{itemize}
\item a given task $\theta_t$ can be assigned at most once to a given worker $\omega_w$, and
\item no more than $r_w$ tasks can be assigned to worker $\omega_w$.
\end{itemize} 
Notice that the second constraint arises from practical considerations on the amount of load a
single worker can tolerate. We also suppose that each single assignment of a task to a worker has a
\emph{cost}, which is independent of the worker's class.  In practical {\color{black} microtask-based} crowdsourcing systems, such cost represents
the {\color{black} low} wages per task the requester pays the worker, in order to obtain answers to his
queries{\color{black}\footnote{We suppose that is the requester has no possibility of refusing the payment for an executed task, whether successful or not.}}. 
{\color{black}In this work,} we assume {\color{black}the same cost for all workers, although it may} appear more natural to differentiate wages among
different classes, so as to incentivize workers to properly
behave~\cite{econ1,econ2}.  Our choice, however, is mainly driven by the following two
considerations: i) while it would be natural to differentiate wages according to the individual
reputation of workers, when the latter information is sufficiently accurate, it is much more
questionable to differentiate them according to only a collective reputation index, such as
$\pi_{t k}$, especially when workers with significantly different reputation coexist within the same
class; ii) since in this paper our main goal is to analyze the impact on system performance of
a-priori available information about the reputation of workers, we need to compare the performance
of such systems against those of systems where the requester is completely unaware of workers'
reputation, under the same cost model. Finally, we wish to remark that both our problem formulation
and proposed algorithms naturally extend to the case in which costs are class-dependent.

Let an \emph{allocation} be a set\footnote{In the following, sets
    are denoted by calligraphic uppercase letters and families of sets are denoted
    by bold calligraphic uppercase letters. Moreover, vectors and matrices are represented by
    lowercase and uppercase bold letters, respectively. The matrix $\Mm$ whose
    elements are $m_{ij}$ is also denoted by $\Mm=\{m_{ij}\}$.} of assignments
of tasks to workers.  More formally, we can represents a generic allocation with
a set $\Gc$ of pairs $(t,w)$ with $t \in \{1,\cdots, T\}$ and $w \in
\{1,\cdots,W\}$, where every element $(t,w)\in\Gc$ corresponds to an individual
task-worker assignment.  Let $\Oc$ be the complete allocation set, comprising
every possible individual task-worker assignment (in other words $\Oc$ is the
set composed of all the possible $T\cdot W$ pairs $(t,w)$). Of course, by
construction, for any possible allocation $\Gc$, we have that $\Gc \subseteq
\Oc$. Hence, the set of all possible allocations corresponds to the power set of
$\Oc$, denoted as $2^{\Oc}$.

The set $\Gc$ can also be seen as the edge set of a bipartite graph where the
two node subsets represent tasks and workers, and there is an edge connecting
task node $t$ and worker node $w$ if and only if $(t,w) \in \Gc$. It will be
sometimes useful in the following to identify the allocation with the
biadjacency matrix of such graph.  Such binary matrix of size $T
  \times W$ will be denoted $\Gm(\Gc) = \{g_{tw}\}$, $g_{tw}\in\{0,1\}$ and referred to as the
  \emph{allocation matrix}.

In this work, we suppose that the allocation is non-adaptive, in the sense that
all assignments are made before any decision is attempted. With this hypothesis,
the requester must decide the allocation only on the basis of the a-priori
knowledge on worker classes. {\color{black} Because of this one-shot assumption, both individual and class error probabilities are considered to be constant over time, as well as constant is the mapping between workers and classes.} Adaptive allocation strategies can be devised as
well, in which, after a partial allocation, a decision stage is performed, and
gives, as a subproduct, refined a-posteriori information both on tasks and on
workers' accuracy. This information can then be used to optimize further
assignments. However, in \cite{shah2} it was shown that non-adaptive allocations
are order optimal in a single-class scenario.

When all the workers' answers are collected, the requester starts deciding,
using the received information. Let $\Am(\Gc) = \{a_{tw}\}$ be a
  $T \times W$ random matrix containing the workers' answers and having the same
  sparsity pattern as $\Gm(\Gc)$. Precisely, $a_{tw}$ is nonzero if and only if
$g_{tw}$ is nonzero, in which case $a_{tw} = \tau_t$ with probability $1-p_{tw}$
and $a_{tw} = -\tau_t$ with probability $p_{tw}$. For every instance of the
matrix $\Am(\Gc)$ the output of the decision phase is an estimate vector
$\hat{\tauv}(\Gc)=[\hat{\tau}_1, \hat{\tau}_2, \dots, \hat{\tau}_T]$ for task
values. In the following, for notation simplicity we will drop the
  dependence of the matrices $\Gm$ and $\Am$ and of the estimates $\hat{\tauv}$
  on the allocation set $\Gc$, except when needed for clarity of presentation.
  
{\color{black} As a final justification of the model described in this section, we describe here a \textit{modus operandi} for a microtask-based crowdsourcing system like MTurk, that would be well modeled by our assumptions. Suppose the platform first calls for a prequalification with respect to a given set of tasks. After the due number of workers have applied, this first phase is closed, and the crowd of potential  workers is formed. In the second phase, such crowd is partitioned into classes according to reputation, and actual task assignment to (a subset of) applicants takes place. Finally, answers are collected and decisions are taken.}

\section{Problem Formulation\label{sec:PF}}
In this section, we formulate the problem of the optimal allocation of tasks to
workers, with different possible performance objectives. We formalize such
problem under the assumption that each worker in class $\Cc_k$ has error
probability for task $\theta_t$ deterministically equal to $\pi_{tk}$.  By
sorting the columns (workers) of the allocation matrix $\Gm$, we can partition
it as 
\begin{equation}
\Gm = \left[ \Gm_1, \Gm_2, \dots, \Gm_K \right] 
\end{equation}
where $\Gm_k$ is a binary matrix of size $T \times W_k$ representing the
allocation of tasks to class-$k$ workers. 

We define $d_{tk}$ as the weight (number of ones) in the $t$-th row of matrix
$\Gm_k$, which represents the number of times task $t$ is assigned to class-$k$
workers. Such weights can be grouped into the $T\times K$ matrix of integers $\Dm(\Gc)
= \{d_{tk}\}$.\\

\noindent {\bf Remark.}  If the individual error probability of the workers
within one class is not known to the scheduler, it becomes irrelevant which
worker in a given class is assigned the task. What only matters is actually how
many workers of each class is assigned each task.  Under this condition
\begin{enumerate}
\item any performance parameter to be optimized can be expressed as a function
  of the weight matrix $\Dm$;
\item any two allocation sets $\Gc_1$ and $\Gc_2$ such that $\Dm(\Gc_1)=\Dm(\Gc_2)$
show the same performance;
\item by optimizing the weight matrix $\Dm(\Gc)$ we also optimize the set of
  allocations $\Gc$.
\end{enumerate}

\subsection{Optimal allocation}
We formulate the problem of optimal allocation of tasks to workers
  as a combinatorial optimization problem for a maximum overall cost. 
Let $\Phi(\Gc)$ be a given performance parameter to be maximized. We
fix the maximum number of assignments (i.e., the maximum number of
  ones in matrix $\Gm$) to a value $C$, and we seek the best allocation
  $\Gc$ as
\begin{eqnarray}  
\Gc^{\mathrm{opt}} 
 &=& \argmax_{\Gc}\,\, \Phi(\Gc) \non
 &\mbox{s.t.}& 0\mathord{\leq} d_{tk}\mathord{\leq} W_k,\,\,t\mathord{=}1,\ldots,T,\,\,k\mathord{=}1,2,\dots,K, \non 
 && \sum_{t=1}^T d_{tk} \leq \sum_{w=W^{(k-1)}+1}^{W^{(k)}} r_w, \,\,k\mathord{=}1,\dots,K, \non
 && \sum_{t=1}^T \sum_{k=1}^K d_{tk} \leq C 
\label{eq:optimal_allocation}
\end{eqnarray}
where $d_{tk}$ are the (integer) elements of $\Dm(\Gc)$,
$W^{(k)} = \sum_{i=1}^k W_{i}$ and $W^{(0)}=0$. The second constraint
in~\eqref{eq:optimal_allocation} expresses the fact that $d_{tk}$ is the number
of ones in the $t$-th row of $\Gm_k$, the third constraint derives from the
maximum number of tasks a given worker can be assigned, and the last constraint
fixes the maximum overall cost.

Note that it could also be possible to define a dual optimization problem, in which the optimization
aims at the minimum cost, subject to a maximum admissible error probability; this alternative problem
is left for future work.

We now denote by $\Fb$ the family of all feasible
allocations (i.e.  the collection of all the allocations respecting the constraints on the total
cost and the worker loads).  Observe that by construction $\Fb \subseteq 2^{\Oc}$ is composed of all
the allocations $\Gc$ satisfying: i) $|\Gc|\le C$, and ii) $|\Lc(w,\Gc)|\le r_w$ $\forall w$, where
$\Lc(w,\Gc)$ represents the set of individual assignments in $\Gc$ associated to $w$.  
\begin{proposition} \label{prop-Matroid}
The family $\Fb$ forms a Matroid~\cite{Calinescu}. Furthermore, $\Fb$ satisfies the following
property.  Let $\Bb \subseteq \Fb$  be the family of maximal sets in $\Fb$, then
$q= \frac{\max_{\Gc \in \Bb}  |\Gc|}{\min_{\Gc \in \Bb}  |\Gc|}=1$.
\end{proposition}
The proof is reported in the Appendix \ref{app:matroid}.

\subsubsection{Computational  complexity}
the complexity of the above optimal allocation problem heavily depends on the
structure of the objective function $\Phi(\Gc)$ (which when specifying the
dependence on the allocation set $\Gc$ can be rewritten as $\Phi(\Gc)$). As a
general property, observe that necessarily $\Phi(\Gc)$ is monotonic, in the
sense that $\Phi(\Gc_1)\mathord{\le} \Phi(\Gc_2)$ whenever
$\Gc_1\mathord{\subset} \Gc_2$. However, in general, we cannot assume that
$\Phi(\Gc)$ satisfies any other specific property (some possible definitions for
$\Phi(\Gc)$ are given next).  For a general monotonic objective function, the
optimal allocation of tasks to workers can be shown to be NP-hard, since it
includes as a special case the problem of the maximization of a monotonic
submodular function, subject to a uniform Matroid constraint
(see~\cite{Calinescu})\footnote{A set function $f: 2^{\Oc}\to \mathbb{R}^+ $ is
  said to be submodular if: $\forall \Ac,\Bc \in 2^{\Oc}$ we have $f(\Ac \cup
  \Bc) + f(\Ac \cap \Bc) \le f(\Ac) + f(\Bc)$. The problem of the maximization
  of a monotonic submodular function subject to a uniform Matroid constraint
  corresponds to: \{$\max_{|\Ac|\le K} f(\Ac)$ for $K<|\Oc|$ with $f(\cdot)$
  submodular.\}}.  When $\Phi(\Gc)$ is submodular, the optimal allocation
problem falls in the class of problems related to the maximization of a
monotonic submodular function subject to Matroid constraints. For such problems,
it has been proved that a greedy algorithm yields a
1/(1+$q$)-approximation~\cite{Calinescu} (where $q$ is defined as in Proposition
\ref{prop-Matroid}).  In the next subsections, we consider different choices for
the performance parameter $\Phi(\Gc)$.

\subsection{Average task error probability }

A possible objective of the optimization, which is most closely related to
typical performance measures in practical crowd work systems, is the average
task error probability, which (except for the minus sign, which is due to the
fact that we need to minimize, rather than maximize, error probability) is
defined as:
\begin{equation} 
\Phi_1(\Gc) = -\frac1{T} \sum_{t=1}^{T} P_{e,t}
\end{equation}
  where
\begin{equation}
P_{e,t} = \PP \{\hat{\tau}_t \neq \tau_t \} = \PP \{\hat{\tau}_t \neq 1 | \tau_t=1\}
\label{eq:Pet}
\end{equation} 
is the error probability on task $t$ and where the second equality
in~\eqref{eq:Pet} follows from the fact that tasks are uniformly distributed in
$\{\pm 1\}$. Note that the probabilities $P_{e,t}$,
  $t=1,\ldots,T$ depend on the allocation set $\Gc$ through the vector of task
  estimates $\hat{\tauv}$.
  Of course, $P_{e,t}$ can be exactly computed only when the true
workers' error probabilities $p_{tw}$ are available; furthermore it heavily
depends on the adopted decoding scheme. As a consequence, in general, $P_{e,t}$
can only be approximately estimated by the requester by confusing the actual
worker error probability $p_{tw}$ (which is unknown) with the corresponding
average class error probability $\pi_{tk}$.  Assuming a maximum-a-posteriori
(MAP) decoding scheme, namely, $\hat{\tau}_t(\alphav) = \arg \max_{\tau_t \in
  \pm 1} \PP \{\tau_t |\av_{t}= \alphav \}$, where $\av_t$ is the $t$-th row of
$\Am$ and $\alphav$ is its observed value, we have
\begin{equation}
\label{eq:error_probability_task_i} P_{e,t} = \sum_{\alphav: \PP \{\tau_t
  =1|\av_{t}= \alphav \} < 1/2} \PP \{\av_{t} = \alphav|\tau_t=1 \}\,.  
\end{equation} 
 It is easy to verify
  that the exact computation of this average task error probability
  estimate requires a number of operations growing exponentially with the number
  of classes $K$.  Thus, when the number of classes $K$ is large, the evaluation
  of (\ref{eq:error_probability_task_i}) can become critical.

To overcome this problem, we can compare the performance of different
allocations on the basis of a simple pessimistic estimate of the error
probability, obtained by applying the Chernoff bound to the random variable that
is driving the maximum-a-posteriori (MAP) decoding (details on a MAP decoding
scheme are provided in the next section). We have:
\[
P_{e,t} \le \widehat{P}_{e,t}=\exp\left(- \frac{ \sum_k d_{tk}(1- 2 \pi_{tk})z_{tk}}{ \sum_k (d_{tk} z_{tk})^2}\right)
\]
where $z_{tk}=\log(\frac{1-\pi_{tk}}{ \pi_{tk}})$. Thus, the performance metric associated with an allocation becomes $\Phi_2(\Gc)= -\frac1{T} \sum_{t=1}^{T} \widehat{P}_{e,t}$.
The computation of $\Phi_2(\Gc)$ requires a number of operations that scales
linearly with the product $T\cdot K$.  However, in practical cases, we expect
the number of classes to be sufficiently small (order of few units), so that the
evaluation of (\ref{eq:error_probability_task_i}) is not an issue.

\subsection{Overall mutual information\label{sec:mutual_info}}
An alternative information-theoretic choice for $\Phi(\Gc)$ is the mutual information between the
vector of RVs associated with tasks $\tauv = (\tau_1, \tau_2,\dots, \tau_T)$ and the answer matrix
$\Am(\Gc)$, i.e., 
\beq
\label{eq:mutual_info}
\Phi_3(\Gc) = I(\Am; \tauv) = \sum_{t=1}^T I(\av_{t}; \tau_t)\,.  
\eeq 
It is well known that a tight
relation exists between the mutual information and the achievable error probability, so that a
maximization of the former corresponds to a minimization of the latter.  We remark, however, that,
contrary to error probability, mutual information is independent from the adopted decoding
scheme, because it refers to an optimal decoding scheme. This property makes the adoption of 
mutual information as the objective function for the task assignment quite attractive, since it
permits to abstract from the decoding scheme.  The second equality in (\ref{eq:mutual_info}) comes
from the fact that tasks are independent and workers are modeled as BSCs with known error
probabilities, so that answers to a given task do not provide any information about other tasks. By
definition \beq
\label{eq:mutual_info_definition}
I(\av_{t}; \tau_t) = H(\av_{t}) - H(\av_{t} | \tau_t) = H(\tau_t) - H(\tau_t |\av_{t} ) 
\eeq 
where $H(a)$ denotes the entropy of the RV $a$, given by\footnote{$\EE_a$ denotes the expectation with respect to RV $a$.} $H(a) = -\EE_a [\log \PP(a)]$  
and for any two random variables $a,b$, $H(a|b)$ is the conditional entropy defined as
$H(a|b) = -\EE_b\EE_{a|b} [\log \PP(a|b)]$.   
In what follows, we assume perfect knowledge of worker reliabilities, i.e., we assume that each
class-$k$ worker has error probability with respect to task $\tau_t$ exactly equal to $\pi_{tk}$,
remarking that in the more general case, the quantities we obtain by substituting $p_{tw}$ with the
corresponding class average $\pi_{tk}$, can be regarded as computable approximations for the true
uncomputable mutual information.

Since we have modeled all workers as BSCs, each single answer is independent of everything else given the task value, so that
\beq \label{eq:conditional_entropy_A_given_t}
H(\av_{t} | \tau_t) = \sum_{a_{tw} \neq 0} H(a_{tw} | \tau_t) = \sum_{k=1}^K d_{tk} H_b(\pi_{tk}) .
\eeq
where $H_b(p) = -p\log p -(1-p) \log (1-p)$.
For the second equality in \eqref{eq:mutual_info_definition}, $H(\tau_t)=1$ because $\tau_t$ is a
uniform binary RV, and
\begin{eqnarray}
H(\tau_t |\av_{t} ) &=& \sum_{\alphav} \PP \{\av_{t} = \alphav\} H(\tau_t |\av_{t}= \alphav  ) \nonumber \\
&=& \sum_{\alphav} \PP \{\av_{t} = \alphav\} H_b(\PP \{\tau_t =1|\av_{t}= \alphav \}  ) 
\end{eqnarray}
where $\alphav$ runs over all possible values of $\av_{t}$.

By symmetry, for every $\alphav$ such that $\PP \{\tau_t =1|\av_{t}= \alphav \} < \frac1{2} $, there
is $\alphav'$ such that $\PP \{\av_{t} = \alphav'\} = \PP \{\av_{t} = \alphav\} $ and $\PP \{\tau_t
=1|\av_{t}= \alphav' \} = 1 - \PP \{\tau_t =1|\av_{t}= \alphav \}$.  As a consequence, we can write
\begin{eqnarray}
H(\tau_t | \av_{t} ) 
&=& 2 \sum_{\mathclap{\alphav: \PP \{\tau_t\mathord{=}1|\av_t\mathord{=}\alphav\} \mathord{<} \frac{1}{2}}} \PP \{\av_t \mathord{=}\alphav\} H_b(\PP \{\tau_t \mathord{=}1|\av_t\mathord{=}\alphav \}) \nonumber \\
&=&\sum_{\mathclap{\alphav:  \PP \{\tau_t \mathord{=}1|\av_{t}=\alphav \} < \frac{1}{2}}}
\left(\PP\{\av_t \mathord{=}\alphav|\tau_t\mathord{=}1\} +\PP\{\av_t\mathord{=}\alphav|\tau_t\mathord{=}\mathord{-}1 \} \right) \cdot \nonumber \\
& & \quad H_b(\PP \{\tau_t =1|\av_{t}= \alphav \}) 
 \label{eq:conditional_entropy_task_given_answers}
\end{eqnarray}
Notice the relationship of the above expression with
\eqref{eq:error_probability_task_i}. If in
\eqref{eq:conditional_entropy_task_given_answers} we substitute $H_b(\PP
\{\tau_t =1|\av_{t}= \alphav \})$ with $\PP \{\tau_t =1|\av_{t}= \alphav \}$,
thanks to Bayes' rule, we obtain~\eqref{eq:error_probability_task_i}.

{\color{black} An explicit computation of $I(\Am; \tauv)$ can be found in Appendix~\ref{app:mi_computation}.
Like for the task error probability, the number of elementary
operations required to compute $I(\Am; \tauv)$ grows exponentially with the
number of classes $K$.}

An important property that mutual information satisfies is submodularity. This property provides
some guarantees about the performance of the greedy allocation algorithm described in
Section~\ref{sec:greedy}.

\begin{proposition}[Submodularity of the mutual information]
\label{prop-submodularity}
Let $\Gc$ be a generic allocation for task $\theta$ and let
  $\av(\Gc)$ be a random vector of answers for task $\theta$. Then, the mutual
  information $I(\av(\Gc); \tau)$ is a submodular function.
\end{proposition}
\begin{IEEEproof}
The proof is given in the Appendix \ref{app:submodularity}.
\end{IEEEproof}

\subsection{Max-min performance parameters}
The previous optimization objectives represent a sensible choice whenever the
target is to optimize the {\em average} task performance. However, in a number
of cases it can be more appropriate to optimize the worst performance among all
tasks, thus adopting a max-min optimization approach.
Along the same lines used in the definition of the previous optimization
objectives, we can obtain three other possible choices of performance parameters
to be used in the optimization problem defined in \eqref{eq:optimal_allocation},
namely, the maximum task error probability, $\Phi_4(\Gc) = -\max_{t=1,\dots,T} P_{e,t}$ the Chernoff bound on the maximum task error
probability, $\Phi_5(\Gc) = - \max_{t=1,\dots,T} \widehat{P}_{e,t}$ and the minimum mutual information,
$\Phi_6(\Gc) = \min_{t=1, 2, \dots, T}I(\av_{t}; \tau_t)$.

\section{Allocation strategies}
\label{sec:allocation}

As we observed in Section \ref{sec:PF}, the optimization problem stated in
\eqref{eq:optimal_allocation} is NP-hard, but the submodularity of the mutual information objective
function over a Matroid, coupled with a greedy algorithm yields a 1/2-approximation~\cite{Calinescu}
(see Proposition \ref{prop-Matroid}). We thus define in this section a greedy task assignment
algorithm, to be coupled with the MAP decision rule which is discussed in the next section.

\subsection{Greedy task assignment\label{sec:greedy}}
The task assignment we propose to approximate the optimal performance is a simple greedy algorithm
that starts from an empty assignment ($\Gc^{(0)}= \emptyset$), and at every iteration $i$ adds to
$\Gc^{(i-1)}$ the individual assignment $(t,w)^{(i)}$, so as to maximize the objective function. In
other words;
\[
(t,w)^{(i)}= \argmax_{\substack{(t,w) \in \Oc\setminus \Gc^{(i-1)}\\(\Gc^{(i-1)}\cup \{(t,w)\})\in \Fb}} \Phi(\Gc^{(i-1)}\cup\{(t,w)\})
\]
The algorithm stops when no assignment can be further added to $\Gc$ without violating some
constraint.

To execute this greedy algorithm, at step $i$, for every task
$t$, we need to i) find, if any, the best performing worker to which task $t$ can be assigned
without violating constraints, and mark the assignment $(t,w)$ as a candidate assignment; ii)
evaluate for every candidate assignment the performance index $\Phi(\Gc^{(i-1)}\cup(t,w))$ $\forall
t$; iii) choose among all the candidate assignments the one that greedily optimizes performance.

Observe that, as a result, the computational complexity of our algorithm is $O(T^2\cdot W Z)$ where
$Z$ represents the number of operations needed to evaluate $\Phi(\Gc)$.

Note that in light of both Propositions~\ref{prop-Matroid} and~\ref{prop-submodularity}, the above
greedy task assignment algorithm provides a $1/2$-approximation when the objective function
$\Phi_3(\Gc)$ (i.e., mutual information) is chosen.  Furthermore, we wish to mention that a better $(1-1/e)$-approximation can be
obtained by cascading the above greedy algorithm with the special local search optimization
algorithm proposed in~\cite{Calinescu}; unfortunately, the corresponding cost in terms of
computational complexity is rather severe, because the number of operations requested to run the
local search procedure is $\widetilde{O}((T\cdot W)^8Z)$.\footnote{The function $f(n)$ is $\widetilde{O}(g(n))$ if $f(n) =O(g(n)\log^bn)$ for any positive
  constant $b$.}

\subsection{Uniform allocation}
Here we briefly recall that \cite{Devavrat,shah2} proposed a simple task allocation strategy (under
the assumption that workers are indistinguishable) according to which a random regular bipartite
graph is established between tasks and selected workers. Every selected worker is assigned the
same maximal number of tasks, i.e. $r_w=r$, $\forall w$, except for rounding effects induced by the
constraint on the maximum total number of possible assignments $C$.

\section{Decision algorithms\label{sec:decision}}

\subsection{Majority voting}
Majority voting is the simplest possible task-decision rule and is currently implemented in all
real-world crowd work systems.  For every task $\theta_t$, it simply consists in counting the
$\{+1\}$ and the $\{-1\}$ in $\av_t$ and then taking $\hat{\tau}_t(\av_t)$ in accordance to the
answer majority.  More formally: 
\beq \label{eq:Majority_decision_rule} 
\hat{\tau}_t(\av_t) = \mathrm{sgn} \left( \sum_w a_{tw} \right).  
\eeq
{\color{black} Note that when $\sum_w a_{tw}=0$, $\hat{\tau}_t(\av_t)$ is randomly chosen in $\{-1,+1\}$.}

\subsection{MAP decision for known workers' reputation}
For the case where each class-$k$ worker has error probability with respect to
task $\tau_t$ deterministically equal to $\pi_{tk}$, the optimal MAP decision
rule can be derived analytically. Indeed, given an observed value of $\av_t$,
the posterior log-likelihood ratio (LLR) for task $\tau_t$ is
\begin{eqnarray}
\mathrm{LLR}_t(\av_t) & = & \log \frac{\PP \{\tau_t = 1 | \av_t\}}{\PP \{\tau_t = -1 | \av_t\}} \non
& = & \sum_{w: a_{tw} \neq 0} \log \frac{\PP \{a_{tw} | \tau_t = 1\} }{\PP \{a_{tw} | \tau_t = -1\} } 
\end{eqnarray}
where the second equality comes from Bayes' rule and the fact that tasks are uniformly distributed
over $\pm 1$, and the third equality comes from modeling workers as BSCs.
Let $m_{tk}$ be the number of ``$-1$'' answers to task $t$ from class-$k$ workers. Then
\beq \label{eq:LLRaposteriori}
\mathrm{LLR}_t(\av_t) = \sum_{k=1}^K \left(d_{tk} - 2 m_{tk}\right) \log \frac{1-\pi_{tk}}{\pi_{tk}}.
\eeq

The MAP decision rule outputs the task solution estimate $\hat{\tau}_t=1$ if $\mathrm{LLR}_t > 0$
and $\hat{\tau}_t=-1$ if $\mathrm{LLR}_t < 0$, that is, \beq \label{eq:MAP_decision_rule}
\hat{\tau}_t(\av_t) = \mathrm{sgn} \left( \mathrm{LLR}_t(\av_t) \right).  \eeq

Observe that the computation of \eqref{eq:LLRaposteriori} has a complexity growing only linearly
with $K$, and that, according to \eqref{eq:MAP_decision_rule}, each task solution is estimated
separately.  Note also that, whenever worker reputation is \emph{not} known a-priori, the above
decision rule is no more optimal, since it neglects the information that answers to other tasks can
provide about worker reputation.

\subsection{ Oracle-aided MAP decision}

The oracle-aided MAP decision rule  is a non-implementable decision strategy which has direct access  to the error probabilities  $p_{tw}$ of every individual worker for every task.

According to the oracle-aided MAP decision rule, first  we compute for every task $\theta_t$:  
\beq \label{eq:OLLRaposteriori}
\mathrm{OLLR}_t(\av_t) = \sum_{w}  a_{tw} \log \frac{1-p_{tw}}{p_{tw}}.
\eeq
Then, the  oracle-aided MAP decision rule outputs the task solution estimate $\hat{\tau}_t=1$ if $\mathrm{OLLR}_t > 0$
and $\hat{\tau}_t=-1$ if $\mathrm{OLLR}_t < 0$, that is, 
\beq \label{eq:OMAP_decision_rule}
\hat{\tau}_t(\av_t) = \mathrm{sgn} \left( \mathrm{OLLR}_t(\av_t) \right).  
\eeq

Observe that the oracle-aided MAP decision rule provides an upper bound to the performance of every implementable  decision rule 
(i.e., it gives a lower bound on the error probability). 

\subsection{Low-rank approximation (LRA)} \label{sec:LRA}
For the sake of comparison, we briefly recall here the Low-Rank Approximation
decision rule proposed in~\cite{Devavrat,shah2,moderators} for the case when: i)
no a-priori information about the reputation of workers is available, ii) the
error probability of every individual worker $w$ is the same for every task,
i.e., $p_{tw}=p_w$ $\forall t$.  The LRA decision rule was shown to provide
asymptotically optimal performance under assumptions i) and ii)~\cite{shah2}.

Denote with $\vv$ the leading right singular vector of $\Am$, the LRA decision
is taken according to:
\[   \hat{\tau}_t(\av_t)=\sgn\left(\mathrm{LRA}(\av_t)\right) \]
where
\[ \mathrm{LRA}(\av_t)=  \sum_w a_{tw} v_w \]
The idea underlying the LRA decision rule is that each component of the leading
singular vector of $\Am$, measuring the degree of coherence among the answers
provided by the corresponding worker, represents a good estimate of the worker's
reputation.

\subsection{Message passing}
Another possible suboptimal decision algorithm is based on message passing
(MP). The fundamental idea behind MP is that, if the allocation matrix $\Gm$ is
sparse, the MAP algorithm can be well approximated by exchanging locally
computed messages, between the nodes of the bipartite graph whose biadjacency
matrix is $\Gm$, for a certain number of iterations. The algorithm is based on
the hypothesis that a given worker behaves in the same way for all tasks, i.e.,
$p_{tw} = p_w$ for all $t$ and $w$, so that $\pi_{tk} = \pi_k$ for all $t$ and
$k$.

Our MP algorithm is an extension of the one described in \cite{shah2}, where we
take into account the a-priori information about worker classes. {\color{black}
  In~\cite{shah2} it is shown that a particular MP algorithm can be seen as an
  efficient implementation of the LRA decision rule. In such MP algorithm,
  workers are seen as a statistical mixture of a hammer ($p_w=0$) and a
  malicious worker ($p_w=1$). Initially, the statistical mixture assigns the
  same weight to both hypotheses, while at each iteration the weights are
  corrected, strengthening or weakening the hammer hypothesis for each worker
  according to whether she has answered correctly most tasks or not. Our
  implementation, instead, assumes a different statistical mixture for each
  class, and a more complex weight update rule, which is essentially locally
  optimal. The details of our implementation follow.  }

In the considered bipartite graph, nodes are either task nodes or worker nodes. An edge connects task node $t$ to worker node $w$ if and only if $g_{tw}=1$. Messages are exchanged along the edges. Let $k(w)$ be the class which worker $w$ belongs to and $f_{k(w)}^{(0)}(p)$ be the a-priori pdf of the error probability for  class $k(w)$. Let $m_{t \rightarrow w}^{(l)}$ and  $m_{w \rightarrow t}^{(l)}$ be the messages sent from task node $t$ to worker node $w$ (resp. from worker node $w$ to task node $t$) in the $l$-th iteration, $l=1,2,\dots$
Given the answer matrix $\Am = \{a_{tw}\}$, the MP algorithm reads as follows.

{\bf Initial condition:} 
\beq \label{eq:initial_pi_MP}
p_{tw}^{(0)} = \pi_{k(w)} 
\eeq
For $l=1,2,\dots$

{\bf Output LLR: } 
\beq \label{eq:outputLLR_MP}
\mathrm{LLR}_{t}^{(l)} = \sum_{  w} a_{tw} \log \frac{1-p_{tw}^{(l-1)}}{p_{tw}^{(l-1)}}
\eeq

{\bf Task-to-worker message: }
\beq \label{eq:task_to_worker_MP}
m_{t \rightarrow w}^{(l)} = \sum_{ w' \neq w} a_{tw'} \log \frac{1-p_{tw'}^{(l-1)}}{p_{tw'}^{(l-1)}}
\eeq

{\bf Worker-to-task message: } 
\beq \label{eq:worker_to_task_MP}
m_{w \rightarrow t}^{(l)} = p_{tw}^{(l)} = \int_0^1 p f_{tw}^{(l)}(p) dp, 
\eeq
being
\beq \label{eq:error_prob_pdf_MP}
f_{tw}^{(l)}(p) \mathord{\propto} f_{k(w)}^{(0)}(p) \prod_{t' \neq t} \left[1\mathord{+} (1\mathord{-}2p) a_{t'w} \tanh \left( \frac{m_{t' \rightarrow w}^{(l)}}{2}\right)\right]
\eeq

It can be seen from~\eqref{eq:outputLLR_MP} that, at each iteration, task nodes
perform the LLR computation as in the MAP decision rule for known workers'
reputation, similarly to \eqref{eq:LLRaposteriori}, with the current estimates
of workers' error probabilities. Because of the initialization in
\eqref{eq:initial_pi_MP}, the LLR outputs at the first iteration are equal to
\eqref{eq:LLRaposteriori}.
 
The task-to-worker message in \eqref{eq:task_to_worker_MP} is the
\emph{extrinsic} LLR, i.e., the one that does not consider the incoming message
on the same edge. The worker-to-task message in \eqref{eq:worker_to_task_MP} is
the updated estimate $p_{tw}^{(l)}$ of the error probability $p_{w}$ of worker
$w$. It is computed as the average with respect to the current pdf
$f_{tw}^{(l)}(p)$ for task $t$ of $p_{w}$, given by
\eqref{eq:error_prob_pdf_MP}. The details of the derivation of
\eqref{eq:error_prob_pdf_MP} are given in Appendix \ref{app:MP_equation}.

Regarding the a-priori pdf, several choices are possible. In our
implementation, we have considered as a-priori pdf $f_k^{(0)}(p)$ the max-entropy
distribution~\cite[Ch. 12]{Cover} over $[0,1/2]$ with a mean equal to $\pi_k$, namely
\beq \label{eq:max_entropy_pdf} f_k^{(0)}(p) \propto \ee^{\lambda_k p} \eeq
where $\lambda_k$ is a parameter that depends on the mean $\pi_k$.
If Haldane priors are assumed for all workers, i.e., 
\beq \label{eq:haldane_pdf}
f_{k}^{(0)}(p) = \frac1{2} \delta(p) + \frac1{2} \delta(p-1)
\eeq
where $\delta(\cdot)$ denotes Dirac delta function, then we obtain the
simplified MP algorithm whose description can be found in~\cite{shah2}. 
Simulation results will show in many cases the advantage of using
\eqref{eq:max_entropy_pdf} instead of \eqref{eq:haldane_pdf}, whose performance
is essentially the same as the LRA decision rule of Section \ref{sec:LRA}.

\section{Results\label{sec:results}}
In this section, except stated otherwise, we evaluate the performance of a
system where $T=100$ tasks are assigned to a set of workers, which are organized
in $K=3$ classes.  Each worker can handle up to 20 tasks, i.e., $r_w=20$,
$w=1,\ldots, W$.
We compare the performance of the allocation algorithms and decision rules described
in Sections~\ref{sec:allocation} and~\ref{sec:decision}, in terms of achieved average error
probability, $P_e=\frac{1}{T}\sum_t P_{e,t}$.  We study the performance of:
\begin{itemize}
\item the ``Majority voting'' decision rule applied to the ``Uniform allocation'' strategy, hereinafter
  referred to as ``Majority uniform'';
\item the ``Majority voting'' decision rule applied to the ``Greedy allocation'' strategy, hereinafter
  referred to as ``Majority greedy'';
\item the ``Low rank approximation'' decision rule applied to the ``Uniform allocation'' strategy,
  in the figures referred to as ``LRA uniform'';
\item the ``Low rank approximation'' decision rule applied to the ``Greedy allocation'' strategy, in the
  figures referred to as ``LRA greedy'';
\item the ``MAP'' decision rule applied to the ``Greedy allocation'' strategy, in the following referred
  to as ``MAP greedy''.
 \item the ``Message Passing'' decision algorithm applied to the ``Greedy allocation'' strategy, in the following referred to as ``MP greedy''. 
\item the ``Oracle-aided MAP'' decision rule applied to the ``Greedy allocation'' strategy, in the following referred
  to as ``OMAP greedy''.
\end{itemize}
Specifically, for the greedy allocation algorithm, described in
Section~\ref{sec:greedy}, we employed the overall mutual information
$\Phi_3(\Gc)$ as objective function.  We consider 4 scenarios characterized by
different number of workers, number of classes, and workers' error
probabilities. The main system parameters for all scenarios are summarized in Table~\ref{table-param}.

\begin{table}
\begin{center}
\caption{Main parameters for the three considered scenarios \label{table-param}}
\begin{tabular}{|l|l|lll|lll|} 
\hline
\hline
                       & $T$  & $\pi_{t1}$ & $\pi_{t2}$ & $\pi_{t3}$ & $W_1$ & $W_2$ & $W_3$ \\
\hline
\hline
              Scenario 1 &100  & 0.1 & 0.2 & 0.5 & 30 &120& 150 \\
\hline            
\multirow{2}{*}{Scenario 2} & 50  & 0.05 & 0.1 & 0.5 & \multirow{2}{*}{30} & \multirow{2}{*}{120}&  \multirow{2}{*}{150} \\
                      & 50  & 0.1 & 0.2 & 0.5 & & &   \\
\hline
\multirow{2}{*}{Scenario 3}  & 50  & 0.1 & 0.2 & 0.5 &  \multirow{2}{*}{40} & \multirow{2}{*}{120}&  \multirow{2}{*}{40} \\
                           & 50  & 0.5 & 0.2 & 0.1 &   & &  \\
\hline
              Scenario 4 &100  & \multicolumn{3}{|c|}{{\small $\pi_{tk}=\frac{2k-1}{4K}$, $1\mathord{\le}k\mathord{\le} K$}} & \multicolumn{3}{|c|}{$\sum_k W_k = 90$} \\
\hline
\end{tabular}
\end{center}
\end{table}

\subsection{Scenario 1}
\begin{figure*}[t]
\centering
\subfigure[]{\includegraphics[width=0.33\textwidth]{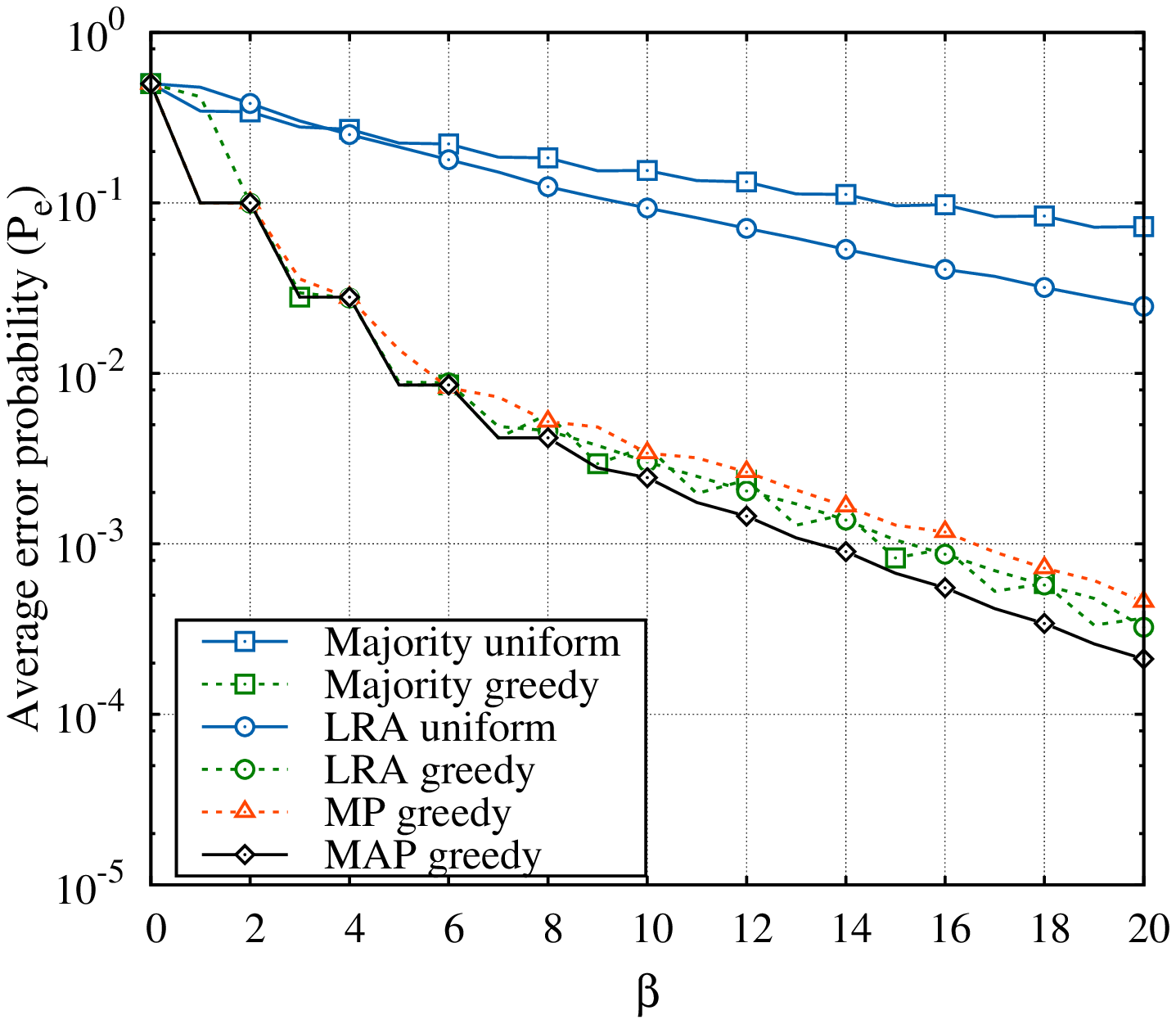}}\hspace*{-5pt}
\subfigure[]{\includegraphics[width=0.33\textwidth]{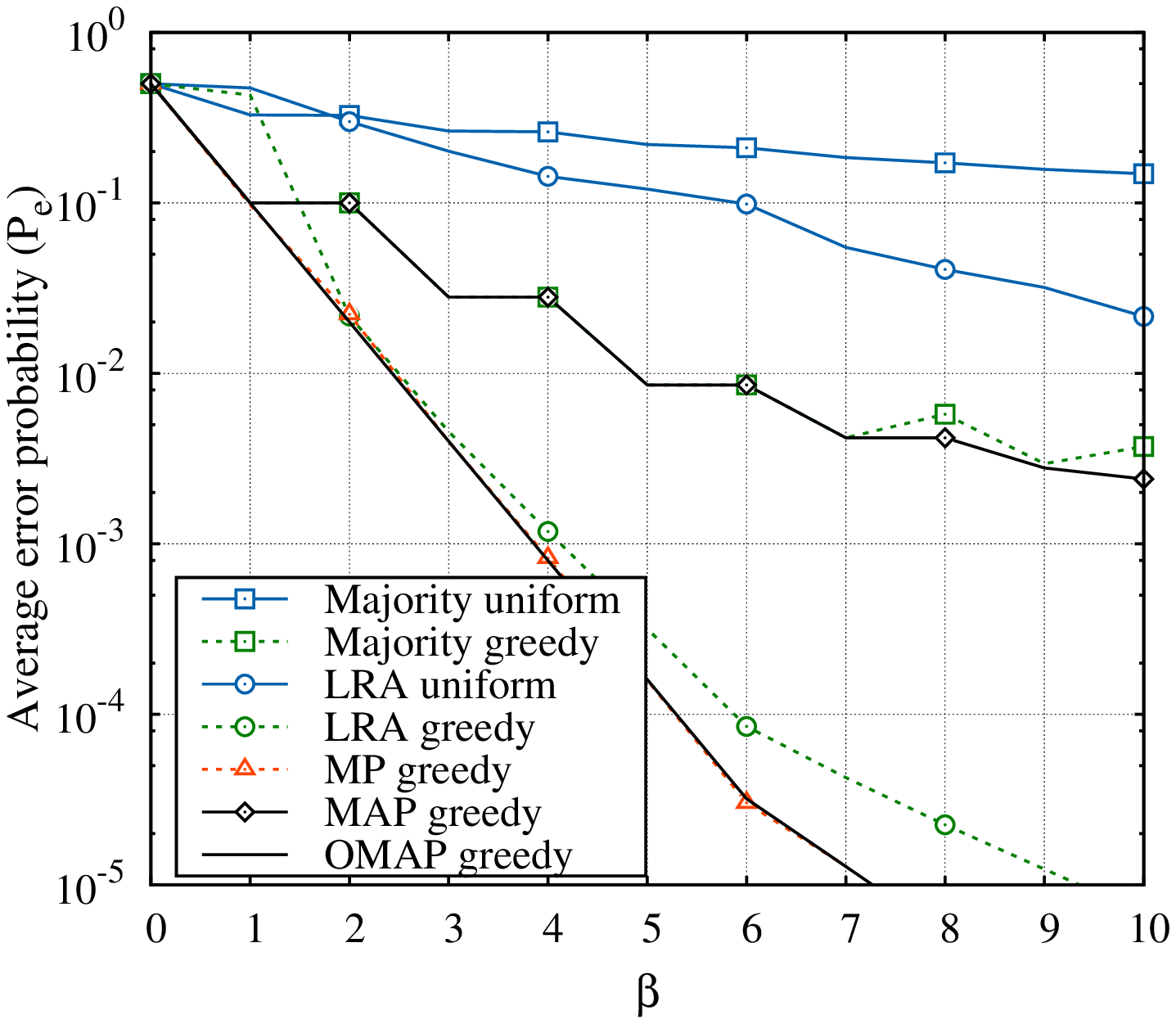}}\hspace*{-5pt}
\subfigure[]{\includegraphics[width=0.33\textwidth]{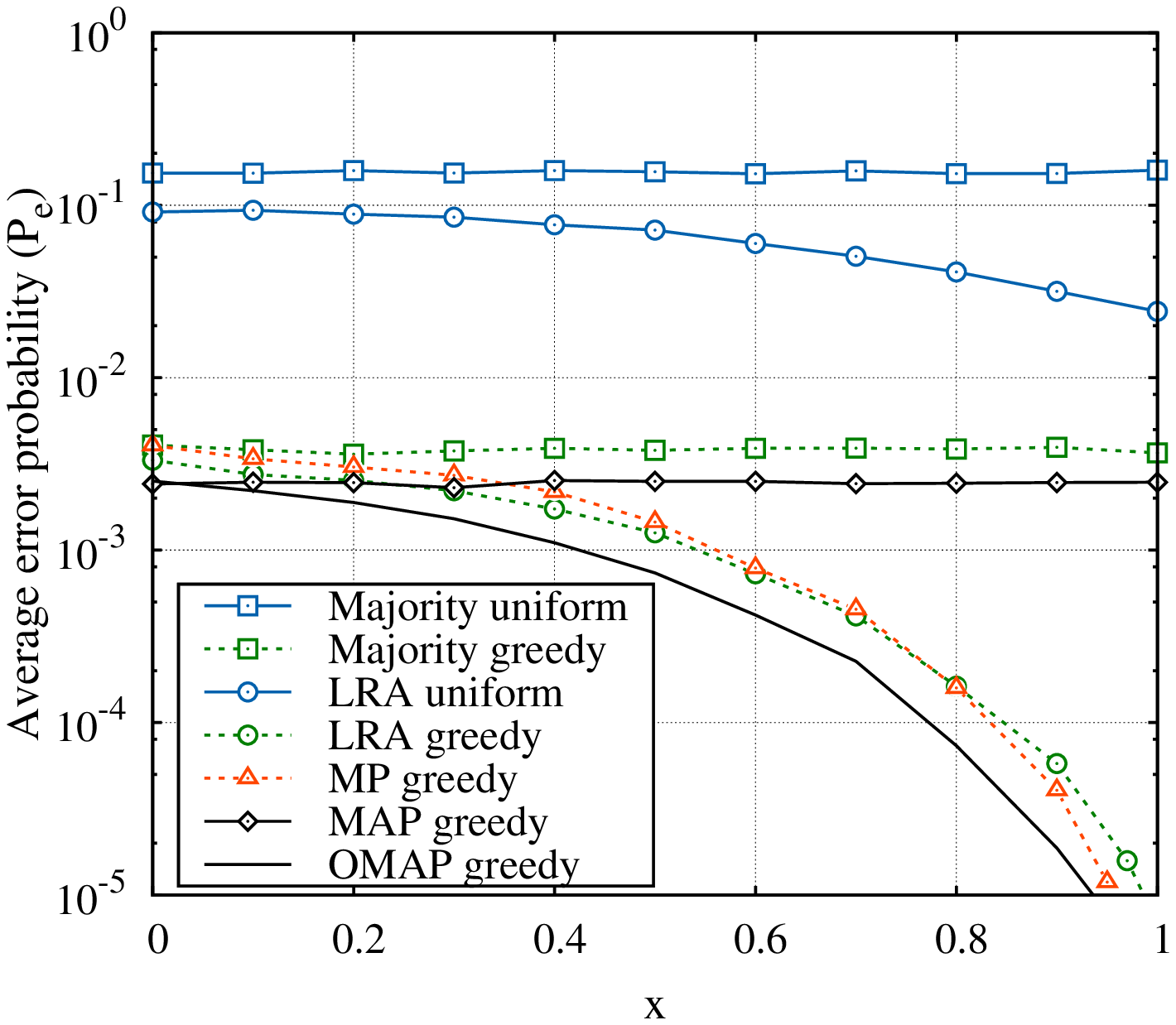}}\hspace*{-5pt}
\caption{Scenario 1. Figures (a) and (b) report the average error probability versus the average number of workers
  per task, $\beta$, for $x=0$ and $x=1$, respectively. Figure (c) shows the average error probability versus $x$ and for
  $\beta=10$. The parameters of Scenario 1 are reported in Table~\ref{table-param}.}
\label{fig:set1}
\end{figure*}

The first set of results is reported in
Figures~\ref{fig:set1}(a),~\ref{fig:set1}(b), ~\ref{fig:set1}(c),
and~\ref{fig:figure1d}.  For starters, results in these figures refer to the
most classical scenario, where all tasks are identical.  For what concerns
workers, we define three classes, and the number of available workers per class
is set to $W_1 = 30, W_2 = 120, W_3 = 150$. {\color{black} Since each worker can
  handle up to 20 tasks ($r_w=20$), the maximum number of assignments that can be handled
  by the three classes are 600, 2400, and 3000, respectively. }

 We set $\pi_{t1}=0.1, \pi_{t2}=0.2, \pi_{t3}=0.5$ for all $t$. This means that
 workers in class 1 are the most reliable, while workers in class 3 are
 spammers.  This situation is summarized in Table \ref{table-param} as Scenario
 1.

The results depicted in Figure~\ref{fig:set1}(a) assume that all workers
belonging to the same class have the same error probability i.e.,
$p_{tw}=\pi_{t,k(w)}$, $k(w)$ being the class worker $w$ belongs to. In
particular, the figure shows the average task error probability, plotted versus
the average number of workers per task, $\beta=C/T$.  As expected, greedy
allocation strategies perform much better due to the fact that they exploit the
knowledge about the workers' reliability ($p_{tw}$), and thus they assign tasks
to the best possible performing workers.  These strategies provide quite a
significant reduction of the error probability, for a given number of workers
per task, or a reduction in the number of assignments required to achieve a
fixed target error probability.  For example, $P_e=10^{-2}$ can be achieved by
greedy algorithms by assigning only 6 workers per task (on average), while
algorithms unaware of workers reliability require significantly more than 20
workers per task (on average).  

{\color{black} Since class 1 can handle up to 600 assignments, for the greedy
  allocation algorithm, and for $\beta>6$ the requester has to allocate tasks to
  workers of class 2 as well. Since workers of class 2 are less reliable than
  workers of class 1 the slopes of the curves decrease. This behavior is not
  shown by algorithms employing uniform allocation since they choose workers
  irrespectively of the class they belong to.}

We also observe that: i) in this scenario the
simple MAP decision rule is optimal (it perfectly coincides with the OMAP) ii)
every decision rule (even the simple majority rule) when applied in conjunction
with a greedy scheduler provides comparable performance with respect to the
optimal MAP algorithm (within a factor 2), iii) the performance of the MP greedy
algorithm is slightly worse than LRA greedy in this scenario; iv) some of the
algorithms such as the ``Majority greedy'' exhibit a non monotonic behavior with
respect to $\beta$; this is a consequence of the fact that in order to provide
best performance, these schemes require the number of workers assigned to tasks
to be odd.

As final remark, we would like to observe that we have also implemented the MP
algorithm proposed in \cite{shah2}, obtaining in all cases results that are
practically indistinguishable from LRA.

Next, we take into account the case where in each class workers do not behave
exactly the same.  As already observed, this may reflect both possible
inaccuracies/errors in the reconstruction of user profiles, and the fact that
the behavior of workers is not fully predictable, since it may vary over time.
Specifically, we assume that, in each class, two types of workers coexist, each
characterized by a different error probability $p_{tw}$. More precisely, workers
of type 1 have error probability $p_{tw} = (1-x)\pi_{tk}$, while workers of type
2 have error probability probability $p_{tw} = (1-x) \pi_{tk} +x/2$, where $0\le
x\le 1$ is a parameter. Moreover, workers are of type 1 and type 2 with
probability $1-2\pi_{tk}$ and $2\pi_{tk}$, respectively, so that the average
error probability over the workers in class $k$ is $\pi_{tk}$.  {\color{black}This bimodal worker model, even if it may appear somehow
artificial, is attractive for the following two reasons: i) it is simple (it
depends on a single scalar parameter $x$), and ii) it encompasses as particular
cases the two extreme cases of full knowledge and hammer-spammer. 
  We would like to remark that the allocation algorithm is unaware of this
  bimodal behavior of the workers. Indeed, it assumes that all workers within
  class $k$ have the same error probability, $\pi_{tk}$.}

  For $x=0$ all workers in each class behave exactly the same (they all have
  error probability $p_{tw}=\pi_{tk(w)}$); this is the case depicted in
  Figure~\ref{fig:set1}(a). For $x=1$ we recover the hammer-spammer scenario;
  this case is represented in Figure~\ref{fig:set1}(b), where workers are
  spammers with probability $2\pi_{tk}$ and hammers with probability
  $1-2\pi_{tk}$.  Here, strategies employing the greedy allocation still greatly
  outperform schemes employing uniform allocations.  However, differently from
  the previous case, the performance of the MAP decision rule is significantly
  sub-optimal in this scenario, when compared to ``LRA greedy``, and ``MP
  greedy''.  This is due to the following two facts: i) MAP adopts a mismatched
  value of the error probability of individual workers, when $x\neq 0$, ii) MAP
  does not exploit the extra information on individual worker reliability that
  is possible to gather by jointly decoding different tasks.  Observe that the
  performance of ``LRA greedy`` and ``MP greedy' is not significantly different
  from OMAP.  In particular the scheme ``MP greedy'' provides performance
  practically indistinguishable from OMAP in this scenario.

In Figure~\ref{fig:set1}(c), for $\beta=10$, we show the error probability
plotted versus the parameter $x$.  We observe that the performance of the ``MAP
greedy'' strategy is independent on the parameter $x$ while the performance of
``LRA greedy'' and ``MP greedy'' improves as $x$ increases. This effect can be
explained by observing that the ability of distinguishing good performing
workers from bad performing workers within the same class increases as $x$
increases.  Observe also that the error probability achieved by ``LRA greedy''
and ``MP greedy'' is within a factor 2 from OMAP for every $x$. In
particular, observe that for small values of $x$ the simple MAP decision rule
represents a low-cost good performing decision strategy, but, as $x$ increases
(i.e. the degree of heterogeneity increases within each class), more
sophisticated decision rules which jointly decode large groups of tasks are
needed to achieve good performance.  It is worth observing that the performance
of the ``MP greedy'' algorithm becomes very close to OMAP for large values of
$x$.

In light of the previous observations, we can conclude that the a-priori
information about worker reliability can be effectively exploited to improve the
overall performance of the system both at the scheduler level (e.g. greedy vs
uniform schemes) and at the decision rule level (e.g. ``MAP'' vs ``Majority
greedy'' for small $x$ and ``MP greedy'' vs ``LRA greedy'' for large $x$).

\insertfig{0.9}{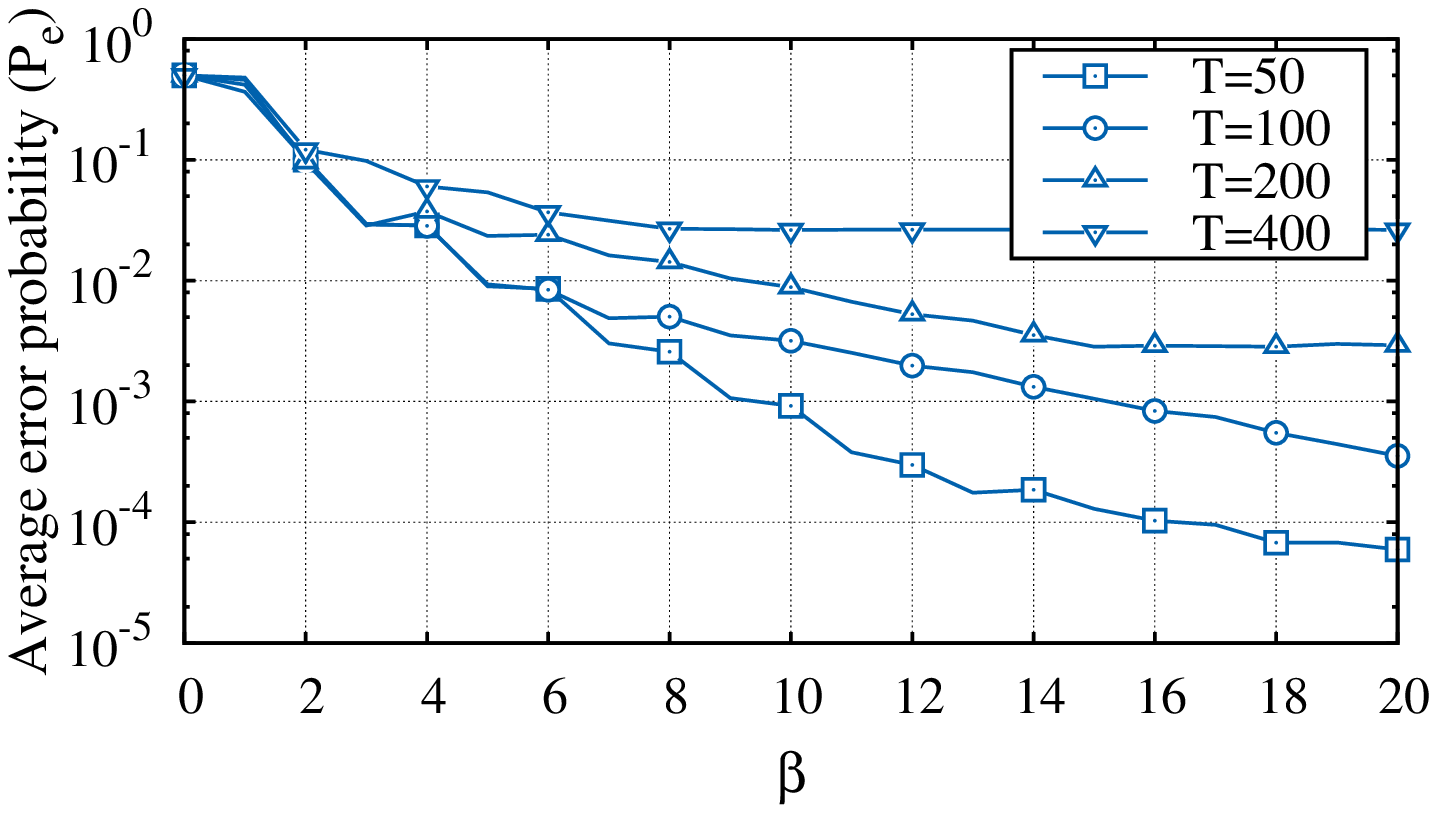}{Average error probability versus $\beta$ for different values of $T$} {fig:figure1d}

{\color{black} In Figure~\ref{fig:figure1d} we show the performance of the ``LRA
  greedy'' algorithm as the number of tasks varies while the pool of workers is
  the same as in the previous figures. Clearly, as the number of tasks increases
  the error probability increases since a larger amount of work is conferred to
  workers of classes 2 and 3. By looking at the curve for $T=200$ we observe
  that (i) for $0\le \beta \le3$ only workers of class 1 are used; (ii) for
  $3<\beta\le 15$ the slope of the curve decreases since workers of both classes
  1 and 2 are used; (iii) for $\beta>15$ the requester allocates tasks also to
  workers of class 3 which are spammers. Therefore for $\beta>15$ the error
  probability does not further decrease. }

\subsection{Scenario 2}
\begin{figure*}[t]
\centering
\subfigure[]{\includegraphics[width=0.33\textwidth]{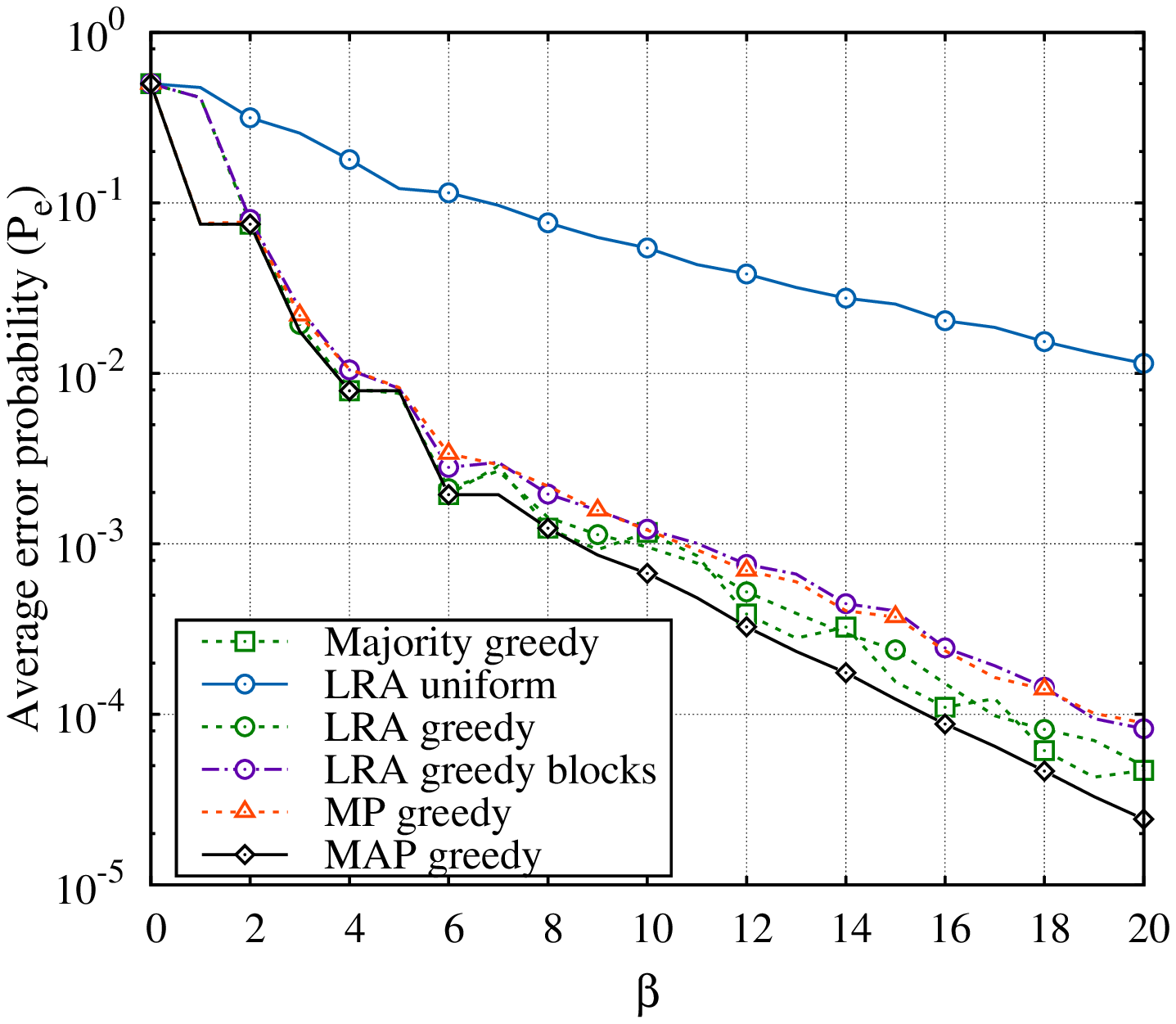}}\hspace*{-5pt}
\subfigure[]{\includegraphics[width=0.33\textwidth]{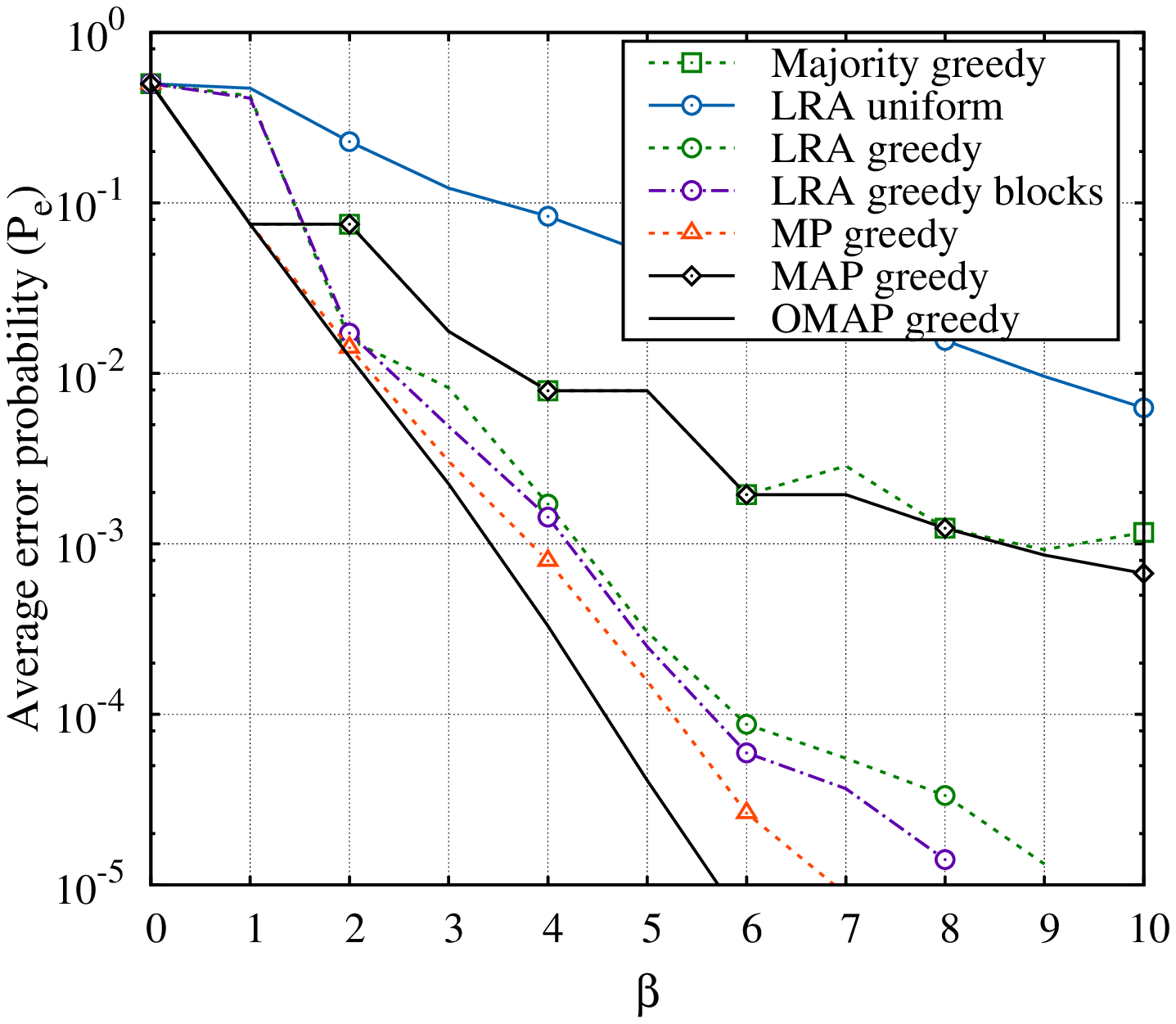}}\hspace*{-5pt}
\subfigure[]{\includegraphics[width=0.33\textwidth]{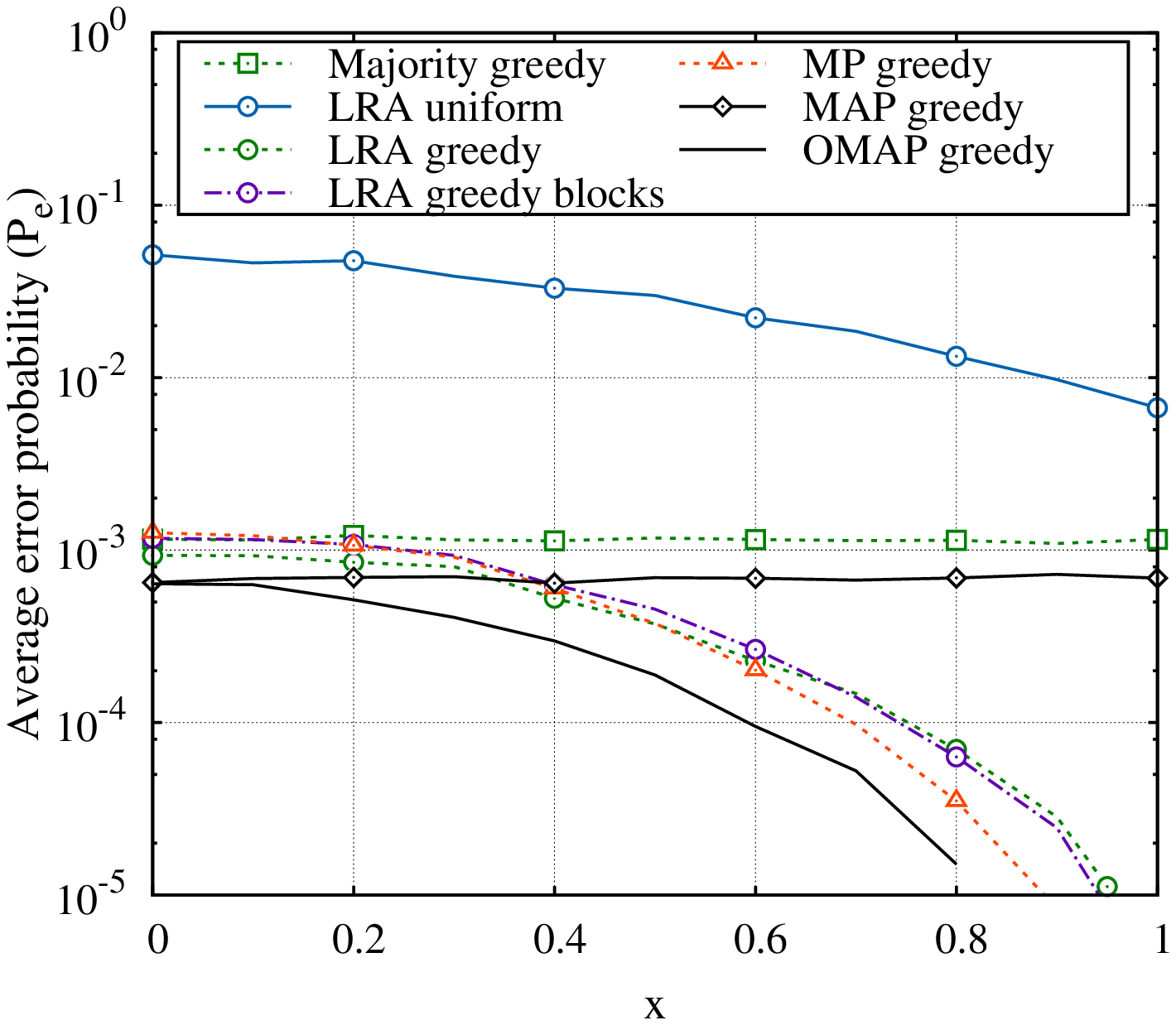}}\hspace*{-5pt}
\caption{Scenario 2. Figures (a) and (b) report the average error probability versus the average number of workers
  per task, $\beta$, for $x=0$ and $x=1$, respectively. Figure (c) shows the average error probability versus $x$ and for
  $\beta=10$. The parameters of Scenario 2 are reported in Table~\ref{table-param}.}
\label{fig:set2}
\end{figure*}
Next, we assume that the $T=100$ tasks are divided into 2 groups of 50
each. Workers processing tasks of group 1 and 2 are characterized by average
error probabilities $\pi_{t1}=0.05, \pi_{t2}=0.1, \pi_{t3}=0.5$ and
$\pi_{t1}=0.1, \pi_{t2}=0.2, \pi_{t3}=0.5$, respectively.  This scenario
reflects the case where tasks of group 2 are more difficult to solve than tasks
of group 1 (error probabilities are higher). Workers of class $\Cc_3$ are
spammers for both kinds of tasks.  This situation is summarized in Table
\ref{table-param} as Scenario 2.
Two different approaches are possible while applying LRA to scenario 2: i)
we can independently apply LRA to blocks of indistinguishable tasks (we denote
this strategy with ``LRA greedy blocks'') placing ourselves on the safer side.
ii) we can embrace a more risky approach by applying directly LRA to the whole
set of tasks (we denote this strategy with ``LRA greedy''). Regarding the MP
decision algorithm, we have applied two independent instances of the algorithm
to the two groups of tasks.
The error probabilities provided by the considered algorithms are plotted in
Figures~\ref{fig:set2}(a), ~\ref{fig:set2}(b), and~\ref{fig:set2}(c). For the
sake of figure readability results of the basic ``Majority uniform'' strategy
are not reported.
First, observe that the relative ranking of all strategies is essentially not
varied with respect to Scenario 1.  In particular, we wish to highlight the
significant performance gain exhibited by strategies employing the greedy
allocation strategy over those employing a uniform allocation, such as ``LRA
uniform``.  Second, observe that at first glance unexpectedly the ``LRA greedy''
slightly outperforms ``LRA greedy blocks'' for small $x$.  This should not be
too surprising, in light of the fact that: i) even if the error probability of
each user depends on the specific task, the relative ranking among workers
remains the same for all tasks, ii) `LRA greedy'' gets advantage from the fact
that all tasks are jointly decoded (i.e. SVD decomposition is applied to a
larger matrix $\Am$ better filtering out noise).

\subsection{Scenario 3}

\begin{figure*}[t]
\centering
\subfigure[]{\includegraphics[width=0.33\textwidth]{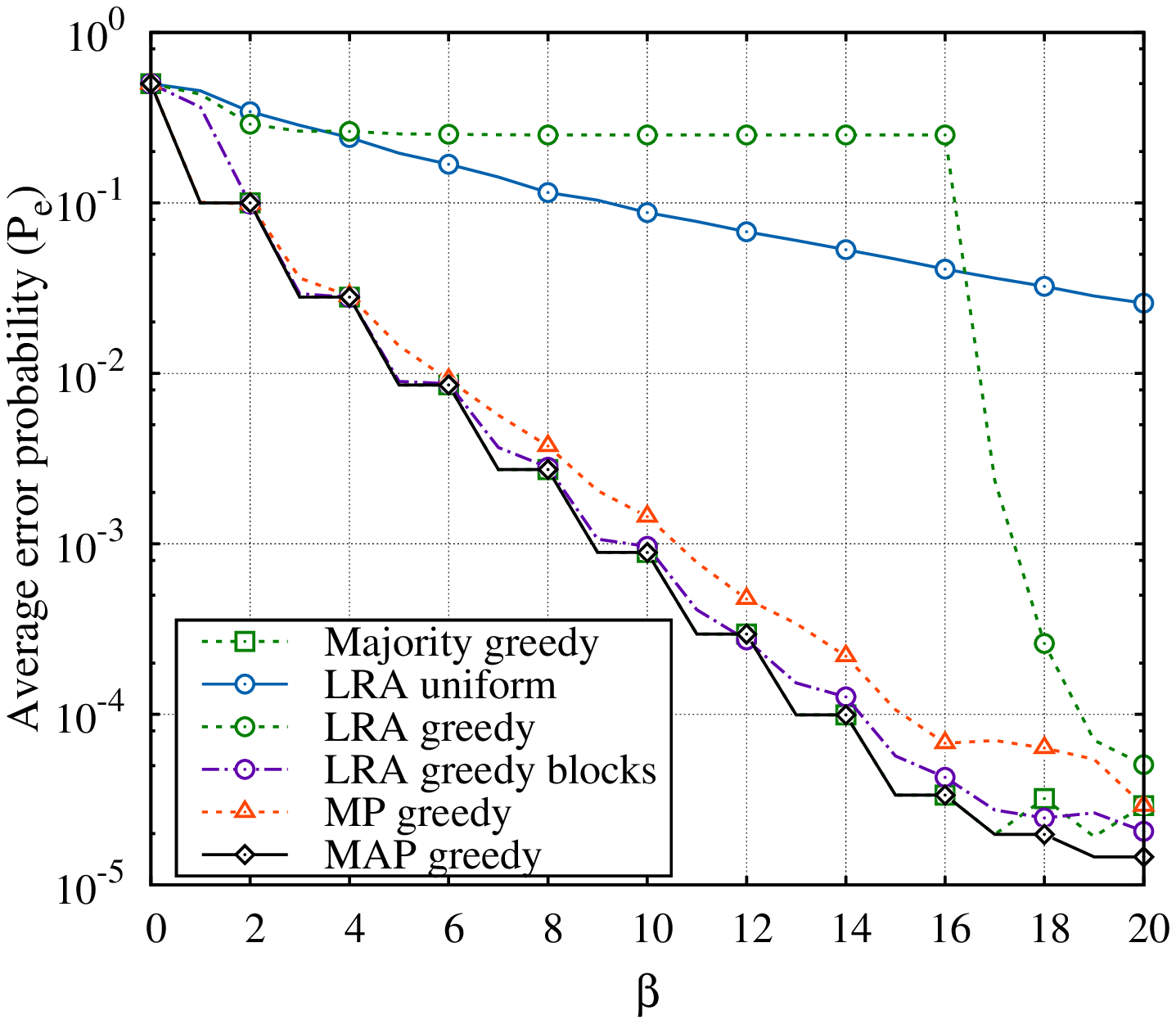}}\hspace*{-5pt}
\subfigure[]{\includegraphics[width=0.33\textwidth]{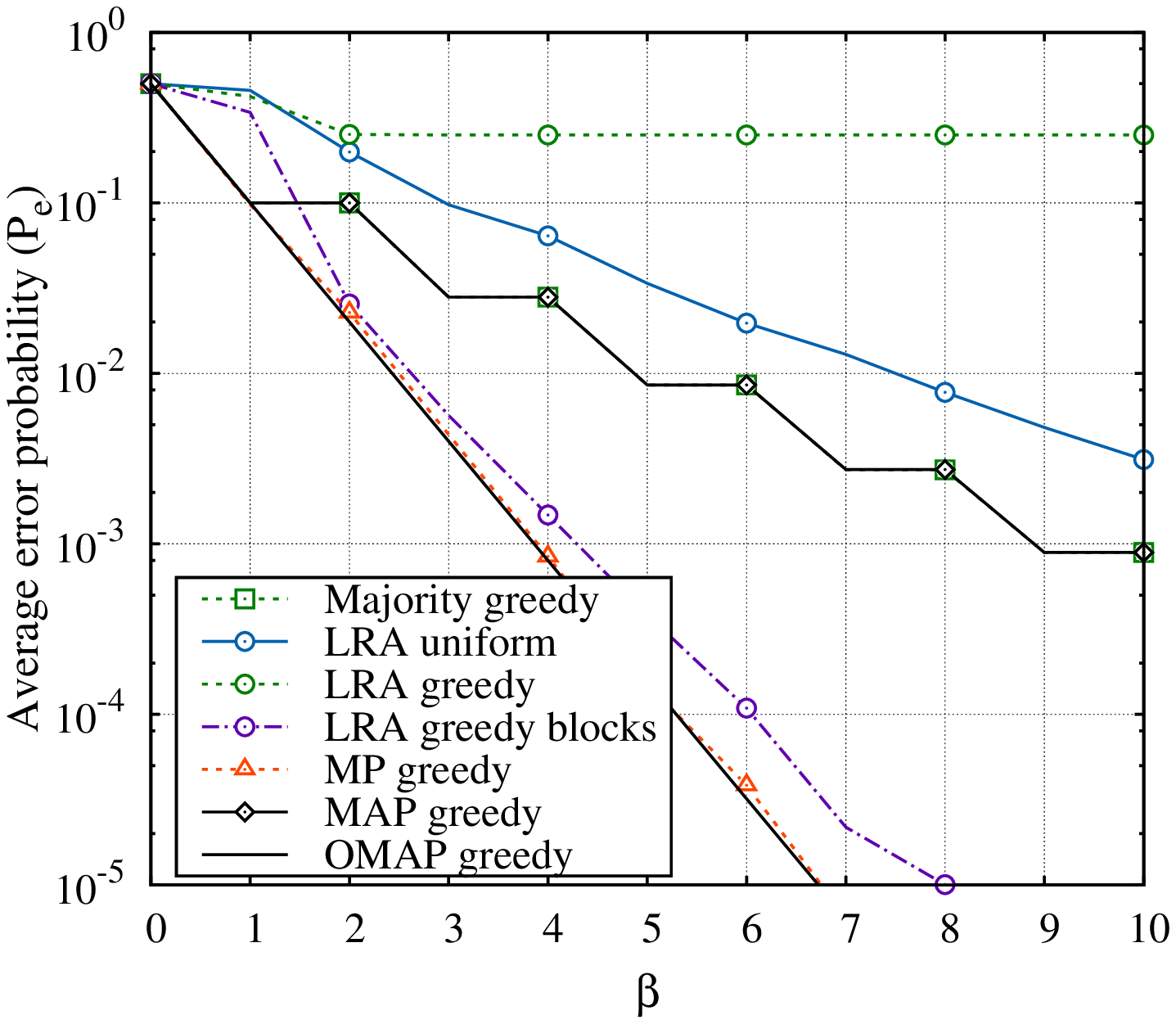}}\hspace*{-5pt}
\subfigure[]{\includegraphics[width=0.33\textwidth]{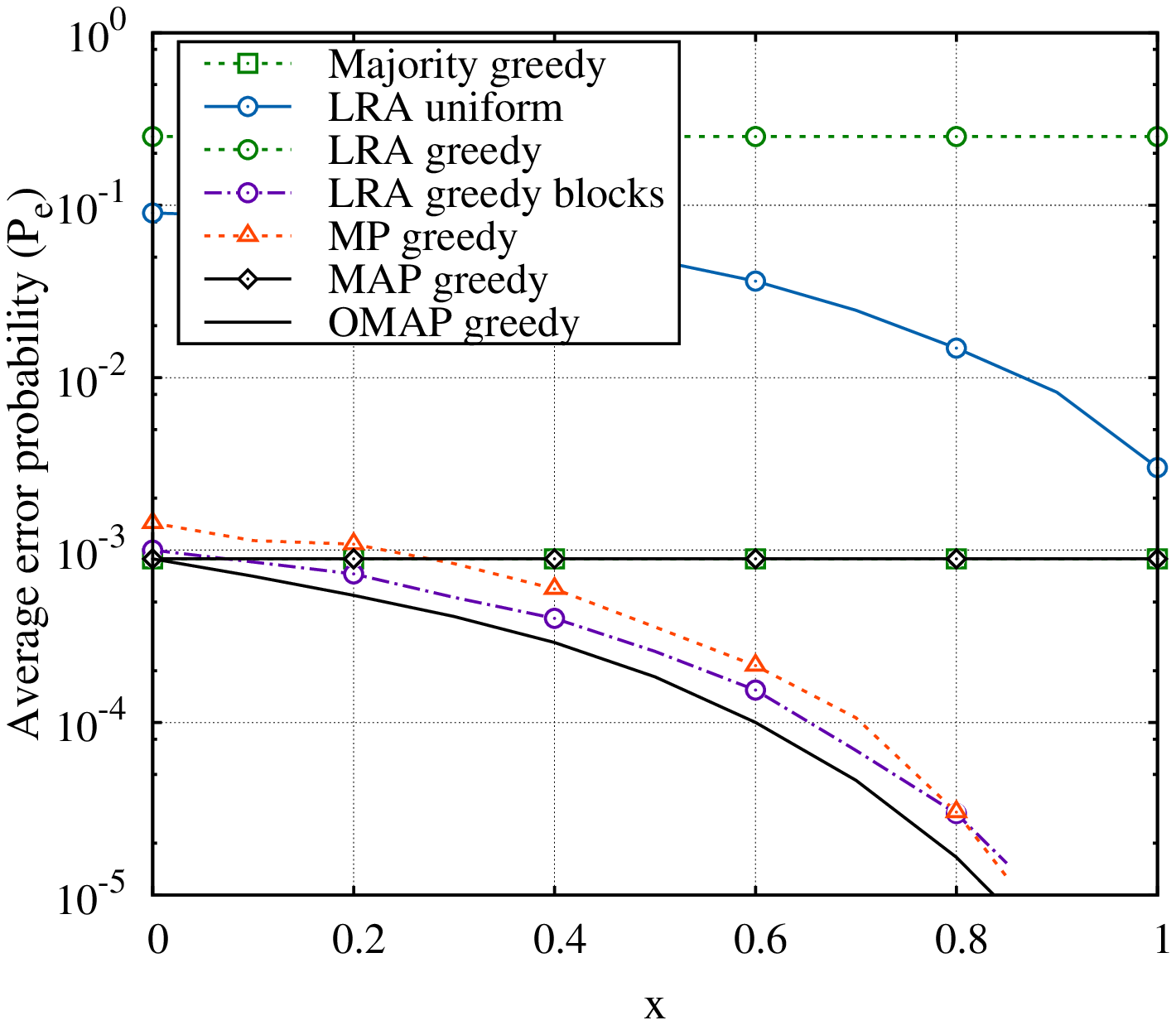}}\hspace*{-5pt}
\caption{Scenario 3. Figures (a) and (b) report the average error probability versus the average number of workers
  per task, $\beta$, for $x=0$ and $x=1$, respectively. Figure (c) shows the average error probability versus $x$ and for
  $\beta=10$. The parameters of Scenario 3 are reported in Table~\ref{table-param}.}
\label{fig:set3}
\end{figure*}

Finally, in Figures~\ref{fig:set3}(a), ~\ref{fig:set3}(b),
and~\ref{fig:set3}(c) we consider a third scenario in which again tasks are
partitioned into two groups of 50 each.  Here, however, the number of available
workers per class is set to $W_1 = 40, W_2 = 120, W_3 = 40$, and the workers
error probabilities for the tasks in group 1 and 2 are given by $\pi_{t1}=0.1,
\pi_{t2}=0.25,\pi_{t3}=0.5$, and $\pi_{t1}=0.5, \pi_{t2}=0.25, \pi_{t3}=0.1$,
respectively. This situation reflects the case where workers are more
specialized or interested in solving some kinds of tasks. More specifically,
here workers of class 1 (class 3) are reliable when processing tasks of group 1
(group 2), and behave as spammers when processing tasks of group 2 (group
1). Workers of class 2 behave the same for all tasks.  This situation is
summarized in Table \ref{table-param} as Scenario 3. We remark that, as in
Scenario 2, we have applied two independent instances of the MP decision
algorithm to the two groups of tasks.
In terms of performance, first observe that, also for this scenario, even a
fairly imprecise characterization of the worker behavior can be effectively
exploited by the requester to significantly improve system performance.  Second
observe that the ``LRA greedy'' algorithm shows severely degraded error
probabilities for $\beta \le 16$, while the ``LRA greedy blocks'' (which we
recall applies LRA independently to the two blocks of indistinguishable tasks)
behaves properly. 
The behavior of ``LRA greedy'' should not surprise the reader, since our third
scenario may be considered as possibly adversarial for the LRA scheme (when
applied carelessly), in light of the fact that the relative ranking among
workers heavily depends on the specific task.  Nevertheless, it may still appear
amazing that ``LRA greedy'' behaves even worse than the ``LRA uniform'' scheme
in several cases.  
{\color{black}
The technical reason for this behavior is related to the fact that, in our
example, for $\beta \le 16$, tasks of group 1 (group 2) are allocated to workers
of class 1 (class 3) only, whilst workers of class 2 are not assigned any
task. Instead, for $16<\beta\le 20$ tasks of both types are also allocated to
workers of class 2. This situation is summarized in
Table~\ref{table-assignments} where $d_{tk}^{(i)}$ represents the number of
times a task of type $i$ is assigned to workers of class $k$.
\begin{table}
\begin{center}
\caption{Assignments per task for Scenario 3 as a function of $\beta$\label{table-assignments}}
\begin{tabular}{|c|cc|cc|cc|} 
\hline
\hline  & $d_{t1}^{(1)}$  & $d_{t1}^{(2)}$ & $d_{t2}^{(1)}$  & $d_{t2}^{(2)}$ &$d_{t3}^{(1)}$  & $d_{t3}^{(2)}$ \\
\hline  $\beta\mathord{\le} 16$ & $\beta$  & 0 & 0 & 0 & 0 & $\beta$ \\
\hline  $16\mathord{<}\beta\mathord{\le} 20$ & $\beta$  & 0 & $\beta\mathord{-}16$ & $\beta\mathord{-}16$ & 0 & $\beta$ \\
\hline
\end{tabular}
\end{center}
\end{table}
For this reason, when $\beta\le16$ the matrix $\Am$ turns out to have a
block diagonal structure, which conflicts with the basic assumption made by LRA
that matrix $\EE [\Am]$ can be well approximated by a rank-1 matrix.}

It follows that the rank-1 approximations are extremely inaccurate when applied
to the whole matrix $\Am$ and thus provide high error probabilities.  In such
situations, by applying the LRA algorithm separately on each block of tasks
(under the assumption that we have enough a-priori information to partitioning
tasks into groups), we achieve huge performance gains.

Finally, we want to remark that we have tested several versions of greedy
algorithms under different objective functions, such as $\Phi_1(\Gc)$,
$\Phi_2(\Gc)$, and $\Phi_3(\Gc)$, finding that they provide, in general,
comparable performance. The version employing mutual information was often
providing slightly better results, especially in the case of LRA greedy and MP
greedy. This can be attributed to the following two facts: i) the mutual
information was proved to be submodular; ii) being mutual information
independent from the adopted decoding scheme, it provides a more reliable metric
for comparing the performance of different task allocations under the LRA
decoding scheme with respect to the error probability $\Phi_1(\Gc)$ (which, we
recall, is computed under the assumption that the decoding scheme is MAP).
Unfortunately, due to the lack of space, we cannot include these results in the
paper.

\subsection{Scenario 4}
 {\color{black} Now, we move to a different scenario in which $T=100$ identical tasks
  are assigned to a population of 90 workers, each one characterized by an
  individual reliability index $p_{t,w}$.  $p_{t,w}$ are uniformly and
  interdependently extracted in the range $(0, 1/2]$.  Parameters $p_{t,w}$ are
assumed, in general, to be unknown to the system, which possesses just noisy
estimates $\hat{p}_{t,w}$ of them.  Such estimates are typically inferred from
the analysis of past behavior of each worker, as better explained in the following.
On the basis of $\hat{p}_{t,w}$, workers are grouped into $K$ classes.
Specifically, classes are obtained by uniformly partitioning the interval $[0,
  1/2]$ in $K$ subintervals and assigning to class $k\in\{1,2,\cdots K\}$ all
the workers whose estimated reliability index (error probability)
$\hat{p}_{t,w}$ falls in the range $(\frac{k-1}{2K}, \frac{k}{2K}]$. Then, the
  nominal error probability $\pi_{tk}$ assigned to class $k$, is set equal to
  the median point of he considered interval, i.e., $\pi_{tk}= \frac{2k-1}{4K}$.

Fig~\ref{fig:figure4a} reports the error probability achieved by the LRA-greedy
algorithm vs $\beta$ for different values of $K$ in a optimistic scenario, in
which perfect estimates of reliability indices are known to the system, i.e.,
$\hat{p}_{t,w}=p_{t,w}$.

As expected, by increasing $K$, a reduction of the error probability is
observed. However note that performance improvements are significant only for
relatively small values of $K$ and $\beta$. The marginal performance gain
observed by increasing $K$ from 6 to 9, is rather limited for all values of
$\beta$. Even in this ideal case in which full information on workers'
characteristics is available to the systems, scheduling tasks just on the basis
of a rough classification of the workers into few classes is not particularly
penalizing!  These results, therefore, provide an empirical justification of
our approach of partitioning users into few classes.

To complement previous results, Fig.~\ref{fig:figure4b} reports the error
probability for specific values of $\beta$ and $K=6$ when the reliability
estimate $\hat{p}_{t,w}$ is noisy.  To keep things simple, we assume that all
workers are tested on an initial number of tasks (called training tasks) and
$\hat{p}_{t,w}$ is derived accordingly, as the empirical estimate of $p_{t,w}$
on the training tasks. However, we wish to remark that $\hat{p}_{t,w}$ can be,
in principle, obtained by analyzing answer of workers on previously assigned
tasks without the necessity of subjecting workers to an initial training phase.
Once again, we would like to highlight that a detailed investigation of how
$\hat{p}_{t,w}$ can be obtained goes beyond the scope of this paper. 

Observe that when the number of training task is 0, no information about
$\hat{p}_{t,w}$ is available and, therefore, workers are assigned to classes at
random.  Instead, when the number of training tasks becomes arbitrarily large,
the system can count on exact estimates of the workers reliability
indices. i.e., $\hat{p}_{t,w}= {p}_{t,w}$.
\insertfig{0.9}{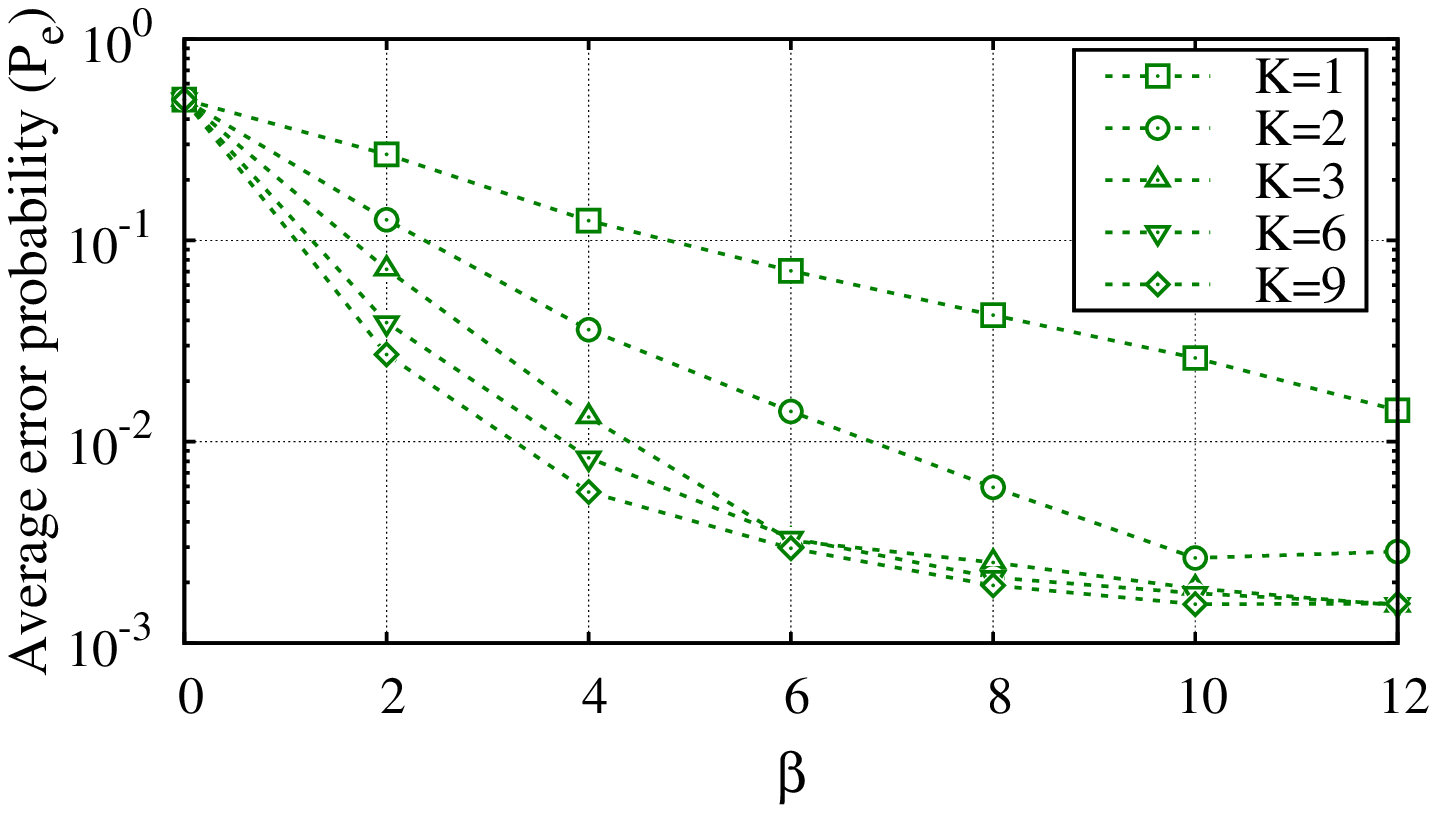}{Average error probability as a
  function of the average number of workers per task, $\beta$, in
  Scenario 4.}{fig:figure4a}
\insertfig{0.9}{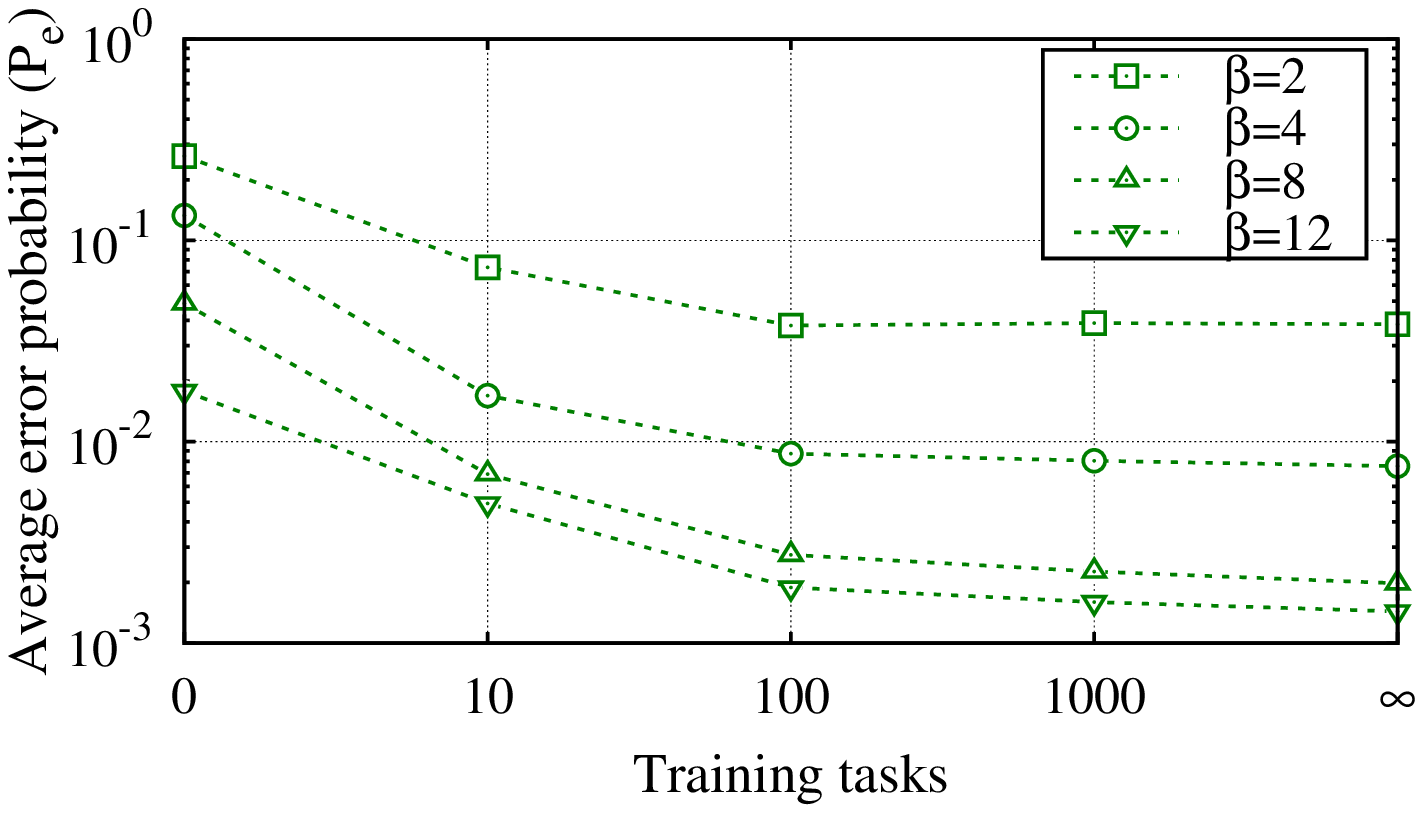}{Average error probability as a
  function of the number of training tasks   in
  Scenario 4.}{fig:figure4b}

Figure ~\ref{fig:figure4b} shows that even rather imprecise estimates of
$\hat{p}_{t,w}$ (i.e., obtained through the analysis of relatively short
sequences of training tasks) can be effectively exploited to significantly
improve the performance of the system.  Furthermore, observe that marginal gains
significantly reduces as the length of training set increases. In particular,
for moderate values of $\beta$, performance obtained when the training set size
is set to 100 is hardly distinguishable from that observed when arbitrarily long
training sets are employed.  This provides further support to the viability of
our approach, which appears rather robust to possibly imprecise estimates of
workers' reliability indices.  }

\section{Concluding Remarks\label{sec:conclusions}}
In this paper we have presented the first systematic investigation of the impact
of information about workers' reputation in the assignment of tasks to workers
in crowd work systems, quantifying the potential performance gains in several
cases.  We have formalized the optimal task assignment problem when workers'
reputation estimates are available, as the maximization of a monotone
(submodular) function subject to Matroid constraints. Then, being the optimal
problem NP-hard, we have proposed a simple but efficient greedy heuristic task
allocation algorithm.  We have also described a simple ``maximum a-posteriori``
decision rule and a well-performing message-passing decision algorithm.  We have
tested our proposed algorithms, and compared them to different solutions, which
can be obtained by extrapolating the proposals for the cases when reputation
information is not available, showing that the crowd work system performance can
greatly benefit from even largely inaccurate estimates of workers' reputation.
Our numerical results have shown that:
\begin{itemize}
 \item even a significantly imperfect characterization of the workers'
   reputation can be extremely useful to improve the system performance;
\item the application of advanced joint task decoding schemes such as message
  passing can further improve the overall system performance, especially in the
  realistic case in which the a-priori information about worker reputation is
  largely affected by errors;
\item the performance of advanced joint tasks decoding schemes such as LRA
  applied naively may become extremely poor in adversarial scenarios.
\item {\color{black} the results show that ``LRA greedy'' and ``MP greedy'' algorithms perform
  well in most of the cases; their difference in terms of performance is rather
  limited, therefore they can both be used equivalently in a real-world
  scenario.}
\end{itemize}

{\color{black} Future work directions include the extension to time-varying
  workers' behavior, non-binary tasks, and the desing of effective algorithms
  for estimating workers' error probability.}

\appendices  

\section{Matroid definition and Proof of Proposition 3.1 \label{app:matroid}}

First we recall the definition of a Matroid. Given a family $\Fb$ of subsets of
a finite ground set $\Oc$ (i.e., $\Fb \subset 2^{\Oc})$, $\Fb$ is a Matroid iff:
i) if $\Gc\in \Fb $, then $\Hc \in \Fb$ whenever $\Hc \subseteq \Gc$;\\ ii) if
$\Gc\in \Fb $ and $\Hc \in \Fb$ with $ |\Gc|>|\Hc|$, then there exists a
$(t_0,w_0) \in \Gc \setminus \Hc$.

Now we can prove Proposition 3.1.  First, observe that in our
case property i) trivially holds. Then, we show that property ii) holds too.
Given that $ |\Gc|>|\Hc|$ and since by construction $ |\Gc| =\sum_w \Lc(w,\Gc)$
and $ |\Hc| =\sum_w \Lc(w,\Hc)$, necessarily there exists an $w_0$ such that
$|\Lc(w_0,\Gc)|>|\Lc(w_0,\Hc)|$. This implies that $\Lc(w_0,\Gc) \setminus
\Lc(w_0,\Hc)\neq \emptyset$.  Let $(t_0, w_0)$ be an individual assignment in
$\Lc(w_0,\Gc) \setminus \Lc(w_0,\Hc)$.  Since by assumption
$|\Lc(w_0,\Hc)|<|\Lc(w_0,\Gc)|\le r_{w_0}$, denoted with $\Hc'= \Hc \cup
\{(t_0,w_0)\}$, we have that $|\Lc(w_0,\Hc')|=|\Lc(w_0,\Hc)| +1 \le
|\Lc(w_0,\Gc)|\le r_{w_0}$ similarly $|\Hc'|=|\Hc|+1\le |\Gc|\le C$, therefore
$\Hc' \in \Fb$.

The fact that in our case $q=\frac{\max_{\Gc in \Bb} |\Gc|}{\min_{\Gc in \Bb}
  |\Gc|}=1$ descends immediately by the fact that necessarily $\Gc \in \Bb$ iff
either i) $|\Gc|=C$ when $C\le \sum_w r_w$ or ii) $ |\Gc|= \sum_w r_w$ when $C>
\sum_w r_w$.

\section{Mutual information for known error probabilities $\pi_{tk}$ \label{app:mi_computation}}
The workers' answers about the tasks $\tauv$ are collected in the random
$T\times W$ matrix $\Am(\Gc)$, defined in Section II of the main document.  The
information that the answers $\Am$ provide about the tasks $\tauv$ is denoted by
\[ I(\Am;\tauv) = H(\Am)- H(\Am|\tauv) \]
where the entropy $H(a)$ and the conditional entropy $H(a|b)$ have been defined in
Section~III.C of the main document.  We first compute $H(\Am|\tv)$ and we observe that, given the tasks
$\tauv$, the answer $\Am$ are independent, i.e., $\PP\{\Am | \tauv\}= \prod_{k=1}^K\prod_{t=1}^T \PP\{\av_{tk}|\tau_t\}$,
where $\av_{tk}$ is the vector of answers to task $\theta_t$ from users of class $\Cc_k$. 
Since $\PP\{\Am | \tauv\}$ has a product form, we obtain $H(\Am|\tauv) = \sum_{k=1}^K\sum_{t=1}^T H(\av_{tk}|\tau_t)$.
Thanks to the fact that workers of the same class are independent and all have error 
probability $\pi_{tk}$, we can write
$H(\av_{tk}| \tau_t) = d_{ik}H_b(\pi_k)$ where $H_b(p) = -p\log p -(1-p)\log(1-p)$ and
$d_{tk}$ is the number of allocations of task $t$ in class $\Cc_k$.  In conclusion, we get:
\[  H(\Am|\tauv) = \sum_{t=1}^T \sum_{k=1}^K  d_{tk} H_b(\pi_{tk})\]

As for the entropy $H(\Am)$, we have:
\[ \PP\{\Am\}  =\EE_{\tauv}\PP\{\Am|\tauv\} = \EE_{\tauv}\prod_{t=1}^T \PP\{\av_t | \tau_t\} = \prod_{t=1}^T \EE_{\tau_t}\PP\{\av_{t} | \tau_t\}\]
 where $\av_t$ is the vector of answers to task $\theta_t$ (corresponding to the $t$-th row of
 $\Am$).  Note that $\EE_{\tau_t}\PP\{\av_{t} | \tau_t\}=\PP\{\av_t\}$, hence $\PP\{\Am\}=\prod_{t=1}^T
 \PP\{\av_t\}$ and we immediately obtain $H(\Am) = \sum_{t=1}^T H(\av_t)$.  The probabilities
 $\PP\{\av_{tk} | \tau_t=1\}$ and $\PP\{\av_{tk} | \tau_i=-1\}$ are easy to compute.  Indeed for
 $\tau_t=-1$ we have
\begin{equation}
  \PP\{\av_{tk} | \tau_t=-1\} = \pi_{tk}^{d_{tk}-m_{tk}}(1-\pi_{tk})^{m_{tk}}
  \label{eq:p_y|t0}
\end{equation}
where $m_{tk}$ is the number of ``$-1$'' answers to task $\theta_t$ from class-$k$ workers.  The above
formula derives from the fact that workers of the same class are independent and have the same
error probability $\pi_{tk}$.  Similarly
\begin{equation}
  \PP\{\av_{tk} | \tau_t=+1\} = \pi_{tk}^{m_{tk}}(1-\pi_{tk})^{d_{tk}-m_{tk}}
  \label{eq:p_y|t1}
\end{equation}
The expressions~\eqref{eq:p_y|t0} and~\eqref{eq:p_y|t1} can compactly written as
\begin{equation}
  \PP\{\av_{tk} | \tau_t\} = (1-\pi_{tk})^{d_{tk}} b_{tk}^{(1-\tau_t)d_{tk}/2-m_{tk}\tau_t} 
  \label{eq:p_y|t}
\end{equation}
where $b_{tk}=\pi_{tk}/(1-\pi_{tk})$. Since, given $\tau_t$, workers are independent, we obtain
\begin{eqnarray}
\PP\{\av_t\}
&=&\EE_{\tau_t}\PP\{\av_{t} | \tau_t\} \non
&=&  \gamma_{tk}\EE_{\tau_t}\left[\prod_{k=1}^K b_{tk}^{(1-\tau_t)d_{tk}/2-m_{tk}\tau_t} \right] \non
&=&  \frac{\gamma_{tk}}{2} \left[\prod_{k=1}^K b_{tk}^{-m_{tk}} + \prod_{k=1}^K b_{tk}^{d_{tk}+m_{tk}} \right] =  \frac{\gamma_{tk}}{2} f(\mv_{t}) \nonumber
\end{eqnarray}
with $\mv_{t} = [m_{t1},\ldots,m_{tK}]$ and $f(\hv) =\prod_{k=1}^K b_{tk}^{-h_{k}} + \prod_{k=1}^K b_{tk}^{d_{tk}+h_{k}}$.
Finally, by using the definition of entropy,
\begin{eqnarray} 
H(\av_t) 
&=& \EE_{\av_t}[-\log \PP\{\av_t\}] \non
&=& -\log\frac{\gamma_{tk}}{2} -\EE_{\av_t}f(\mv_{t}) \non
&=& -\log\frac{\gamma_{tk}}{2} -\frac{\gamma_{tk}}{2}  \sum_{\nv} f(\nv)\log f(\nv)\prod_{k=1}^K\binom{m_{tk}}{n_k}\non \label{eq:Hat}
\end{eqnarray}
where $\nv=[n_{1},\ldots,n_{K}]$ and $n_k=0,\ldots,m_{tk}$, $k=1,\ldots,K$.
Note that the computation of~\eqref{eq:Hat} is exponential in the number of classes $K$.

\section{Submodularity of the mutual information\label{app:submodularity}}
Let $\Gc_1$ and $\Gc_2$ be two generic allocations for task $\theta$, such that
$\Gc_2 \subseteq \Gc_1$ and $\Gc_1 = \Gc_2 \cup \Gc_3$. Also let the pair
$\gamma=(t,w)\in \Oc \setminus \Gc_1$.  Let $\av(\Gc)$ be the random vector of
answers corresponding to the allocation $\Gc$.  For the sake of notation
simplicity in the following we define $\av_j=\av(\Gc_j)$, $j=1,2,3$, and
$a_{\gamma} = \av(\gamma)$ (since $\gamma$ is a single pair, the answer
$\av(\gamma)$ is scalar).

Then the mutual information $I(\av(\Gc); \tau)$ is submodular if
\begin{equation}
I(\av(\Gc_2 \cup \gamma); \tau)-I(\av_2; \tau) \ge I(\av(\Gc_1 \cup \gamma); \tau)-I(\av_1; \tau)\,.
\label{eq:submodularity_I}
\end{equation}
We first observe that
\begin{eqnarray*}
I(\av(\Gc_1 \cup \gamma); \tau) 
&=& I(\av(\Gc_2 \cup \Gc_3 \cup \gamma); \tau) \non
&\stackrel{(a)}{=}& I(\av_2, \av_3,a_{\gamma}; \tau) \non
&\stackrel{(b)}{=}& I(\av_2, a_{\gamma}; \tau) + I(\av_3; \tau | \av_2, a_\gamma)
\end{eqnarray*}
where in $(a)$ we exploited the fact that the sets $\Gc_3$, $\Gc_2$, and
$\gamma$ are disjoint while in $(b)$ we applied the mutual information chain
rule.  Similarly we can write $I(\av_1; \tau) = I(\av_2; \tau) + I(\av_3; \tau |
\av_2)$ By consequence~\eqref{eq:submodularity_I} reduces to
\[ I(\av_3; \tau | \av_2) \ge I(\av_3; \tau | \av_2, a_\gamma) \]
By applying to both sides of the above inequality the definition of the mutual
information given in the main document, equation (7), we obtain
\begin{equation} H(\av_3 | \av_2)\mathord{-}H(\av_3 | \tau, \av_2) \mathord{\ge} H(\av_3 | \av_2, a_\gamma)\mathord{-}H(\av_3 |\tau, \av_2, a_\gamma)
\label{eq:submodularity_I2}
\end{equation}
Since workers are independent we now observe that $H(\av_3 | \tau, \av_2) = H(\av_3 | \tau)$
since $\av_3$ only depends on $\tau_t$. Therefore~\eqref{eq:submodularity_I2} reduces to
\[ H(\av_3 | \av_2) \ge H(\av_3 | \av_2, a_\gamma)  \]
which holds due to the fact that entropy is reduced by conditioning.

\section{Derivation of equation (24) of the main document\label{app:MP_equation}}

In the $l$-th iteration of the MP algorithm, an updated pdf of the error
probability of worker $w$, $p_w$, is computed given the answer matrix $\Am$ and
the current posterior probability distribution of task solutions, used as
a-priori.

Let $\rho_t^{(l)}(\pm 1)$ be the posterior probability distribution for task $t$ at iteration $l$. Also, let $f_{k(w)}^{(0)}(p)$ be the a-priori distribution of the error probability for class $k(w)$ which worker $w$ belongs to. In the computation of the pdf $f_{tw}^{(l)}(p)$ of $p_w$, we use \emph{extrinsic} information, i.e., we only use $\rho_{t'}^{(l)}(\pm 1)$ for $t \neq t'$. We have, thanks to Bayes' rule, that
\beq \label{eq:pr_appD_1}
f_{tw}^{(l)}(p) \propto f_{k(w)}^{(0)}(p) \PP\{\Am | p_{w} = p,  \{\rho_{t'}^{(l)}(\pm 1)\}_{t' \neq t} \}
\eeq
where 
\beq \label{eq:pr_appD_2}
\PP\{\Am | p_{w} \mathord{=} p,  \{\rho_{t'}^{(l)}(\pm 1)\}_{t' \neq t} \} \mathord{\propto} \prod_{t' \neq t}  \PP\{a_{t'w} | p_w \mathord{=} p,  \rho_{t'}^{(l)}(\pm 1) \}
\eeq
In both equations above, the omitted factors do not depend on $p$. Now
\beq \label{eq:pr_appD_3}
\begin{split}
\PP\{a_{tw} | p_{w} = p,  &\rho_{t}^{(l)}(\pm 1) \} = \\
&\PP\{a_{tw} | p_{w} = p,  t=1 \} \rho_{t}^{(l)}( 1) \\
&+ \PP\{a_{tw} | p_{w} = p,  t=-1 \} \rho_{t}^{(l)}( -1)
\end{split}
\eeq
and
\beq \label{eq:pr_appD_4}
\PP\{a_{tw} | p_{w} = p,  t=\pm 1 \} = \frac1{2} \left[ 1 \pm (1-2p) a_{tw}\right]
\eeq
Substituting back \eqref{eq:pr_appD_4} into \eqref{eq:pr_appD_3}, we obtain
\beq \label{eq:pr_appD_5}
\begin{split}
\PP\{a_{tw} | p_{w} = p,  &\rho_{t}^{(l)}(\pm 1) \} \\
&= \frac1{2} \left[ 1 + (1-2p) a_{tw} \left(\rho_{t}^{(l)}( 1) - \rho_{t}^{(l)}(- 1) \right)\right]
\end{split}
\eeq
Finally, from the definition of LLR, we have
\beq \label{eq:pr_appD_5bis}
m_{t \rightarrow w}^{(l)} = \log\frac{\rho_{t}^{(l)}( 1)}{\rho_{t}^{(l)}( -1)}
\eeq
so that
\beq \label{eq:pr_appD_6}
\rho_{t}^{(l)}( 1) = \frac{\ee^{m_{t \rightarrow w}^{(l)}}}{1+\ee^{m_{t \rightarrow w}^{(l)}}}
\eeq
and
\beq \label{eq:pr_appD_7}
\rho_{t}^{(l)}( 1) - \rho_{t}^{(l)}(- 1) =  \tanh \left( \frac{m_{t \rightarrow w}^{(l)}}{2}\right)
\eeq
Substituting \eqref{eq:pr_appD_7} into \eqref{eq:pr_appD_5}, and then back into \eqref{eq:pr_appD_2} and \eqref{eq:pr_appD_1}, 
we obtain equation (24) appearing in the main document.

\end{document}